\documentclass{aa}  

\usepackage{graphicx}
\usepackage{txfonts}
\usepackage{color,soul}
\usepackage{lscape}
\usepackage{caption}
\usepackage{rotating}
\usepackage{float}

\newcommand{\kms}{km s$^{-1}$}
\newcommand{\vlsr}{v$_{lsr}$}

\newcommand{\msol}{M$_{\odot}$}
\newcommand{\lsol}{L$_{\odot}$}
\newcommand{\mysou}{G328.2551-0.5321}

\newcommand{\methanol}{CH$_3$OH}

 \def\as     {\ifmmode {\rlap.}$\,$''$\,$\! \else ${\rlap.}$\,$''$\,$\!$\fi}
\def\decsec  {\ifmmode {\rlap.}$\,$^{\rm s}$\,$\! \else ${\rlap.}$\,$^{\rm s}$\,$\!$\fi}\def\decs  {\ifmmode {\rlap.}$\,$^{\rm s}$\,$\! \else ${\rlap.}$\,$^{\rm s}$\,$\!$\fi}

\defcitealias{Csengeri2018}{Paper I} 

\begin{document}

  % \title{Emergence of complex organic molecules and heavy water in shocks around a young high-mass protostar}
   %\title{The search for high-mass protostars with ALMA revealed up to kilo-parsec scales (SPARKS): II.}
   \title{SPARKS II.: Complex organic molecules in accretion shocks around a hot core precursor}
%   \subtitle{Complex organic molecules and heavy water in shocks around a young high-mass protostar}
% OR: C versus N bearing species in different shocks

%   \subtitle{I. Overviewing the $\kappa$-mechanism}

   \author{T. Csengeri 
          \inst{1,2}
          \and
           A. Belloche
          \inst{1}
           \and
          S. Bontemps
          \inst{2}
          \and
          F. Wyrowski
           \inst{1}
          \and
         K. M. Menten
          \inst{1}
          \and
         L. Bouscasse
          \inst{1}
             }

   \institute{Max Planck Institute for Radioastronomy,
              Auf dem H\"ugel 69, 53121 Bonn, Germany
              \email{csengeri@mpifr-bonn.mpg.de}
                \and
Laboratoire d'astrophysique de Bordeaux, Univ. Bordeaux, CNRS, B18N, all\'ee Geoffroy Saint-Hilaire, 33615 Pessac, France
    }

   \date{Received February 7, 2019; accepted }

% \abstract{}{}{}{}{} 
% 5 {} token are mandatory
 
  \abstract
   {Classical hot cores are rich in molecular emission, and they show a high abundance of complex organic molecules (COMs). The emergence of molecular complexity is poorly constrained in the early evolution of hot cores.
   }
   { We put observational constraints on the physical location of COMs in a resolved  high-mass protostellar envelope associated with the G328.2551$-$0.5321 clump. The protostar is single down to $\sim$400\,au scales and we resolved the envelope structure down to this scale. %on $<$1000\,au scales.  
   }
   {High angular resolution observations using the Atacama Large Millimeter Array allowed us to resolve the structure of the inner envelope and  pin down the emission region of COMs. We use local thermodynamic equilibrium modelling of the available 7.5\,GHz bandwidth around $\sim 345$~GHz to identify the COMs towards two accretion shocks and {\textbf a selected} position representing the bulk emission of the inner envelope. {We quantitatively discuss the derived molecular column densities and abundances towards these positions, and use our line identification to qualitatively compare this to the emission of COMs seen towards the central position, corresponding to the protostar and its accretion disk.}
   }
   {We detect emission from 10 COMs, and identify a line of deuterated water (HDO). In addition to methanol (\methanol), methyl formate (CH$_3$OCHO) and formamide (HC(O)NH$_2$) have the most extended emission. Together with HDO, these molecules are found to be associated with both the accretion shocks and the inner envelope, which has a moderate temperature of $T_{\rm kin}\sim$110\,K. We find a significant difference in the distribution of COMs. O-bearing COMs, such as ethanol, acetone, and ethylene glycol are almost exclusively found and show a higher abundance towards the accretion shocks with $T_{\rm kin}\sim$180\,K. Whereas N-bearing COMs with a CN group, such as vinyl and ethyl cyanide peak on the {central position, thus the protostar and the accretion disk}. 
     The molecular composition is similar towards the two shock positions, while it is significantly different towards the inner envelope, suggesting an increase in abundance of O-bearing COMs towards the accretion shocks. 
   }
   {We present the first observational evidence for a large column density of COMs seen towards accretion shocks at the centrifugal barrier at the inner envelope. The overall molecular emission shows increased molecular abundances of COMs towards the accretion shocks compared to the inner envelope. The  bulk of the gas from the inner envelope is still at a moderate temperature of $T_{\rm kin}\sim$ 110\,K, and we find that the radiatively heated  inner region is very compact {($<$1000\,au). Since the molecular composition is dominated by that of the accretion shocks and the radiatively heated hot inner region is very compact, we propose this source to be a precursor to a classical, radiatively heated hot core. }
   %The extended emission of methyl formate and HDO at this moderate temperature suggests a grain-surface production and subsequent sublimation to the gas phase for these molecules. The higher abundance of saturated nitriles is similar to what is observed in solar system bodies, such as in the atmosphere of Titan, and can be explained by their higher chemical stability.
   By imaging the physical location of HDO, we find that it is consistent with an origin within the moderately heated inner envelope, suggesting that it originates from sublimation of ice from the grain surface {and its destruction in the vicinity of the heating source has not  been efficient yet.} %Its high abundance of up to $2.1\times10^{-7}$ may indicate a high degree of deuterium fractionation.
}
% What does the HDO/H2O ratio suggest towards hot cores?

   \maketitle
%
%-------------------------------------------------------------------

\section{Introduction}
The origin of complex organic molecules (COMs) that emerge during the process of star and planet formation is a key question in understanding our astrochemical origins.  Historically, COMs in the interstellar medium have been identified towards so called hot cores associated with sites of high-mass star formation \citep{Blake1987, Garay1999,Kurtz2000}. 
In these objects, radiative heating from the central protostar leads to an increase in temperature that can reach T$\gtrsim$100$-$200\,K over an extent of 0.05$-$0.1\,pc, leading to a boost of chemical complexity due to thermal desorption of heavier molecules from the ice mantles and grain surfaces, and subsequent gas-phase reactions. As a result, classical hot cores exhibit a high abundance of COMs (e.g.\,\citealp{Bisschop2007, Mookerjea2007, Beltran2009, Weaver2017}).

{The formation of COMs was first modelled by invoking gas-phase chemical reactions (e.g.\,\citealp{Millar1991,Charnley1992, Caselli1993}); recent chemical models assign, however, an important role to reactions on the grain-surface (\citealp{Charnley2001, Garrod2006, Garrod2013}, see however \citealp{Balucani2015}).} According to these models, COMs are thought to form in the ice mantles of interstellar dust grains \citep{HerbstvanDishoeck2009}. First, through the hydrogenation of atoms and small molecules, the 'zeroth-generation' species of small saturated COMs, such as \methanol, form. The diffusion and recombination of these small molecules and radicals lead to the emergence of 'first-generation' COMs, which becomes efficient at $\gtrsim$30\,K \citep{Garrod2006}. When the radiative feedback from the emerging protostar heats up its environment to $\gtrsim$100\,K,  molecules sublimate from the grains and the 'second-generation' COMs form through gas-phase reactions. 

% SCOPE:
{This relatively simple picture of hot core formation and chemistry seems to be, however, more complicated because while all hot cores are rich in molecular emission, they
%While they are all rich in molecular emission, classical hot cores 
exhibit a significant diversity in their chemical composition (\citealp{Walmsley1993,Kurtz2000,Churchwell2002,Bisschop2007,Calcutt2014,Weaver2017,Bonfand2017,Sanchez-Monge2017,Allen2018}). The origin of this diversity is not fully understood. }
To understand the key  processes leading to the formation of COMs, first the physical conditions, in which these molecules appear, need to be constrained. In practice this means determining in which component of the envelope they reside, allowing us to pinpoint whether they originate from the cold or the heated parts of the envelope, or are associated with slow or fast shocks. Constraining the chemical differentiation of the inner envelope observationally is therefore particularly important because resolving the physical location of COMs reveals the physical conditions of the gas, which can help to constrain their chemical formation pathways. This can be used as a tool to investigate both the physical and chemical processes related to star formation and the emergence of molecular complexity \citep{Garrod2013, Sakai2014}. 

% Hystorically: Orion KL \citealp{Blake1996, Wright1996}

Measured with single dish telescopes, incapable of resolving individual massive envelopes, the origin of chemical differentiation is, however, challenging to identify. Reaching the scales of a few hundred au resolution ($<0\rlap{.}{\arcsec}1$ at distances of several kpc) is necessary to resolve the spatial distribution of COMs within high-mass protostellar envelopes. While this information would provide valuable input to constrain chemical models, this is a largely unexplored territory for the precursors of high-mass stars. In the past, due to angular resolution and sensitivity limitations, typically only the brightest hot cores have been studied (e.g.\,\citealp{Palau2011,Jimenez-Serra2012,WidicusWeaver2012,Friedel2012,Oberg2013,Palau2017, Allen2017}).
These are frequently found in regions in which confusion due to the clustered nature of (high-mass) star formation limits the possibility to reveal the spatial location of COMs within single envelopes.

High angular resolution and high sensitivity observations with the Atacama Large Millimeter Array (ALMA) provide now an increasingly rich insight into the close vicinity of forming O-type stars. An emerging number of observational studies show in unprecedented detail the resolved structure of high-mass clusters and protostellar envelopes (\citealp{Sanchez-Monge2013,Maud2017,Ginsburg2017,Csengeri2018}, hereafter Paper I). Some show an increasing complexity towards smaller scales in terms of clustering, while rare examples of single high-mass protostellar envelopes have been identified \citepalias{Csengeri2018}.  
{Here we study the molecular composition of the gas in the immediate vicinity of \mysou, a hot core precursor, on %the 
%inner 
few hundred au  
scales. }
{This source is the only massive object embedded in the MSXDC G328.25-00.51 dark cloud that is located
at a distance of 2.5$^{+1.7}_{-0.5}$\,kpc. It was identified by
 \citet{Csengeri2017a} 
 based on }
 the APEX Telescope Large Area Survey of the Galaxy (ATLASGAL) \citep{schuller2009, Csengeri2014, Csengeri2017a} at 870\,$\mu$m, and was observed in the frame of the SPARKS project (Search for High-mass Protostars with ALMA revealed up to kilo-parsec scales, Csengeri et al., in prep).
In \citetalias{Csengeri2018} we show that the physical structure of this source is dominated by a single collapsing envelope down to $\sim$400\,au scales (Fig.\,\ref{fig:overview}), which makes it an ideal laboratory to study the emergence of hot cores. {We estimate the protostellar mass to be between $\sim$11 and 16\,\msol\ with an envelope mass $\sim$120\,\msol\,\citepalias{Csengeri2018}.} For the first time, we find indication for shocked gas in the 300-800\,au vicinity of the protostar that has been identified using  a rotational transition of \methanol\ from within its first torsionally excited state, and is interpreted to outline accretion shocks {(Fig.\,\ref{fig:overview}, \textsl{A} and \textsl{B} positions, \citetalias{Csengeri2018})}. This phenomenon has been observed towards nearby low-mass protostars \citep{Sakai2014,Oya2017}, and is expected to arise due to the infall from the envelope onto a compact, rotationally supported accretion disk. {Here we focus on the molecular composition of the observed accretion shocks and reveal spatially resolved emission from several COMs as new tracers pinpointing these shocks.}  %A closer inspection reveals a different spatial distribution between O- and N-bearing COMs due to chemical differentiation. We also identify a transition from HDO, and find that its spatial distribution outlines the  inner region of the moderately heated envelope ($T_{\rm gas}\sim100$\,K) within a radius of $\sim$1000\,au. 

The paper is organised as follows: in Section\,\ref{sec:obs} we present the observations and the data reduction; in Section\,\ref{sec:results} we show the spectra towards the selected positions and analyse their molecular composition. In Section\,\ref{sec:discussion} we discuss the results, and finally in Section\,\ref{sec:conclusions} we present our conclusions.

%quenches the development of an HIIregion(Walmsley et al.1995;Keto2002,2003)

%The hot core
%phase is thought to last about 10^5years(van Dishoeck &
%Blake1998)to 10 6 years(Garrod & Herbst2006) and represents the most chemically rich phase of the interstellar medium, characterized by the presence of complex organic molecules.

\section{Observations and data reduction}\label{sec:obs}

The observations have been carried out with ALMA 
in Cycle\,2 using 35 of the 12\,m antennas on 2015 May 3, and 2015 September 1. The phase
centre  was $(\alpha,\delta)_{\rm J2000}=(15^{\rm h}58^{\rm m}00.05^{\rm s}$, $-53^\circ57'57\rlap{.}{''}8)$ and the baseline range is 15\,m (17\,k$\lambda$) to 1574\,m (1809\,k$\lambda$). 
The total time on source was 7.4 minutes, and the $T_{\rm sys}$ varies between 120 and 200\,K. {The calibrators have been J1517$-$2422 (bandpass), J1617$-$5848 (phase) and the absolute flux scale has been calibrated based on observations of Titan and Ceres. The atmospheric conditions have been stable over both measurement sets with a precipitable water vapour of $\sim$0.8 and $\sim$1.1\,mm for the more extended and the compact configuration, respectively. We estimate an absolute flux uncertainty around $\sim$10\%  by comparing the measured and catalogue fluxes for the two measurement sets for the bandpass and phase calibrators. The bandpass calibrator was measured less frequently and shows on average larger flux variations over the timescale of the two measurement sets compared to the phase calibrator, which does not show strong flux variations over this period. Here we measure a maximum discrepancy between the catalogue and the measured value up to 12\%. } 
We used four basebands in Band 7 centred on
347.331, 345.796, 337.061, and 333.900\,GHz, respectively.  
This gives a $4\times1.75$\,GHz
effective bandwidth with a spectral resolution of 0.977\,MHz  corresponding to $\sim$0.9\,\kms\
velocity resolution.

%-------------------------------------------------------------
%                                             Two column Figure 
%-------------------------------------------------------------
   \begin{figure}[!t]
   \centering
         {\includegraphics[width=0.98\linewidth]{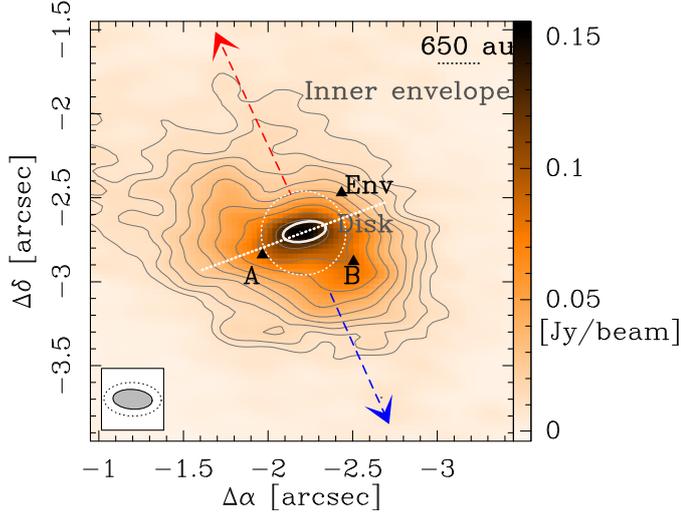}}
      \caption{Line-free continuum emission map at 345\,GHz combining  ALMA 12m and 7m arrays from \citetalias{Csengeri2018} imaged with robust parameter of $-2$ giving a beam size with a geometric mean of $0\as16$. The colour scale is linear between $-3\sigma$ and $120\sigma$, where 1$\sigma$ is 1.3\,mJy/beam,
                      contours start at $7\sigma$ and increase on a logarithmic scale up to $120\sigma$ by a factor of 1.37.  The red and blue dashed lines show the direction of the CO outflow. {The dotted line indicates the direction perpendicular to the outflow, and the white ellipse shows the position of the accretion disk from \citetalias{Csengeri2018}.} The HPBW of the synthesised beam is shown in the lower left corner; the filled ellipse corresponds to that of the continuum image, the dotted ellipse to the molecular line data of this paper. Black triangles mark the positions where the spectra have been extracted for this work, {and the white dotted circle shows the area with a radius of 0\rlap{.}{\arcsec}5 where the spectra have been averaged. }}
         \label{fig:overview}
   \end{figure}
%
%-------------------------------------------------------------

\setlength{\tabcolsep}{1.2pt}
\begin{table}[!ht]
\caption{Summary of detected simple molecules and their transitions.}\label{tab:lines}
\centering
\begin{tabular}{lclrccrrrrrrrr}
\hline\hline
Molecule & Transition &  Frequency & $E_{\rm up}/k$ & Database  \\
&   & [GHz] & [K] & &\\
\hline
$^{34}$SO & 7$_8$ --6$_7$ & 333.901 & 80 & CDMS  \\
SO$_2$ & 8$_{2,6}$--7$_{1,7}$ & 334.673 & 43 & CDMS  \\
H$^{13}$CCCN	 & 38-37  & 334.930 & 314 & CDMS \\
HC$^{13}$CCN?	 & 37-36  & 335.093  & 305 & CDMS \\
HDCO?		 & $5_{1,4}$--$4_{1,3}$  & 335.097  & 56 & CDMS \\
HDO & 3$_{3,1}$--4$_{2,2}$ & 335.396 & 335 & JPL  \\  
SO$_2$ & 29$_{5,25}$--30$_{2,28}$ & 335.773 & 463 & CDMS  \\ % This is quite tentative!!
SO$_2$ & 23$_{3,21}$--23$_{2,22}$ & 336.089 & 276 & CDMS  \\  
HC$_3$N  & 37-36 & 336.521& 307 & CDMS \\
SO &  10$_{11}$--10$_{10}$ & 336.554   & 143 & CDMS \\
SO$_2$ & 16$_{7,9}$--17$_{6,12}$ & 336.670 & 245 & CDMS  \\  
C$^{17}$O &  3--2  & 337.061$^a$ &  32 & CDMS \\
SO$_2$ & 13$_{2,12}$--12$_{1,11}$ & 345.339 & 93 & CDMS  \\  
H$^{13}$CN  & 4--3 & 345.340 & 41 & CDMS \\
$^{34}$SO$_2$ & 6$_{4,2}$--6$_{3,3}$ & 345.553 & 57 & CDMS  \\
HC$_3$N		  & 38--37 & 345.609 & 324 & CDMS \\
$^{34}$SO$_2$ & 5$_{4,2}$--5$_{3,3}$ & 345.651 & 51 & CDMS  \\
CO &  3--2  & 345.796 & 33 & CDMS  \\
NS $\Omega$=1/2, $l$=$e$ &   15/2 - 13/2 & 345.823$^b$ & 70 & CDMS \\
NS $\Omega$=1/2, $l$=$f$ &   15/2 - 13/2 & 346.221$^c$ & 70 & CDMS  \\
HC$_3$N, $\varv_{7}$=1, $l$=1$e$	 & 38-37  & 346.456 & 645 & CDMS  \\
SO$_2$ & 16$_{4,12}$--16$_{3,13}$ & 346.524 & 165 & CDMS  \\  
SO & 9$_8$ -- 8$_7$ & 346.529 & 79 & CDMS  \\
SO$_2$ & 19$_{1,19}$--18$_{0,18}$ & 346.652 & 168 & CDMS  \\  
HC$_3$N, $\varv_{7}$=1, $l$=1$f$	 & 38-37  & 346.949 & 646 & CDMS  \\
H$^{13}$CO$^+$  & 4--3 & 346.998 & 42 & CDMS \\
SiO &  8--7 & 347.331 & 75 & CDMS \\
HN$^{13}$C  & 4--3 & 348.340& 42 & CDMS \\
SO$_2$ & 24$_{2,22}$--23$_{3,21}$ & 348.388 & 292 & CDMS  \\  
NS $\Omega$=3/2  &   15/2 - 13/2 & 348.516$^c$ & 390 & CDMS \\
H$_2$CS & $10_{1,9}$--$9_{1,8}$ & 348.534 & 105 & CDMS \\
CH$_3$CN & 19 --18, K=8$^d$ & 349.025 & 624 & CDMS \\% & CDMS &\\
\hline

CH$_3$OH$-A$ $\varv_{\rm t}=0$    & $2_{-2}-3_{-1}$  & 335.134   & 45 &   CDMS \\
CH$_3$OH$-A$ $\varv_{\rm t}=0$  & $7_1-6_1$  & 335.582  &79&  CDMS\\
CH$_3$OH$-A$ $\varv_{\rm t}=0$   & $14_7-15_6$  & 336.438 &  488 &      CDMS \\
CH$_3$OH$-A$ $\varv_{\rm t}=0$    & $12_{-1}-12_{0}$  & 336.865  &   197 &      CDMS \\
CH$_3$OH$-E$ $\varv_{\rm t}=0$    & $3_3-4_2$ & 337.136   &   62 &    CDMS \\
CH$_3$OH$-E$ $\varv_{\rm t}=1$ & $3_{0}$--$2_{1}$ & 334.427  &  315 &  CDMS \\ 
CH$_3$OH$-A$ $\varv_{\rm t}=2$    & $7_1-6_1$  & 336.606 &  747 &  JPL  \\
$^{13}$CH$_3$OH$-A$ $\varv_{\rm t}=0$ & $12_{-1}$--$12_{0}$ & 335.560  &   193  & CDMS \\
$^{13}$CH$_3$OH$-A$ $\varv_{\rm t}=0$ & $14_{-1}$--$14_{0}$  & 347.188 &    254  & CDMS\\
\hline

\end{tabular}\\
\tablefoot{\\
\tablefoottext{a} {The frequency corresponds to the $F$=1/2-3/2 transition.} \\
\tablefoottext{b} {The frequency corresponds to the $F$=17/2-15/2 transition.} \\
\tablefoottext{c} {The frequency corresponds to the $F$=15/2-13/2 and  $F$=13/2-11/2 transitions.} \\
\tablefoottext{d} {The K=9 transition at 348.911 GHz is likely to be detected, although it is blended with a CH$_3$OCHO line.} \\
}
\end{table}

The data 
have been calibrated in CASA 4.3.1 with the pipeline (version 34044). 
We imaged and cleaned the data using Briggs weighting
and used the CLEAN algorithm for deconvolution. In this work 
we used a robust parameter of $0.5$ for the imaging\footnote{For the line-free continuum maps presented in \citetalias{Csengeri2018} we used a robust parameter of $-2$ to give more weight to the longer baselines yielding a better angular resolution of $0\rlap{.}{\arcsec}16$ corresponding to $\sim$400\,au physical scales.} to favour sensitivity. %, which gives a 
%synthesised beam of 0.34\arcsec$\times0.20$\arcsec\ corresponding, on average, to a $0.26$\arcsec\ resolution ($\sim$650\,au)

We first created a cube of the entire bandwidth with a synthesised half-power beam width (HPBW) of $0\rlap{.}{\arcsec}30$$\times$$0\rlap{.}{\arcsec}17$ and a position angle of 88.37 degrees. The geometric mean of the major and minor axes corresponds to a beam size of $0\rlap{.}{\arcsec}23$ ($\sim$575\,au). 
We then extracted spectra of the entire frequency coverage towards representative positions corresponding to the two accretion shock spots, and a position representing the inner envelope (Fig.\,\ref{fig:overview}). The spectra were then exported to GILDAS/CLASS\footnote{http:www.iram.fr/GILDAS/} for further processing. We subtracted the continuum emission from these spectra by selecting line emission free channels, and using a zero order baseline. To convert the spectra from Jy/beam to Kelvin units we used a factor of 198 based on the geometric mean of the synthesized HPBW.

We also show here images of molecular line emission, where we imaged a narrow velocity range around the selected line. In an iterative process, we first created the image, then identified the emission free channels towards the continuum peak and close to the selected line emission. We then  subtracted the continuum emission in the $uv$-data, performed the gridding and the deconvolution procedure, as described above, to obtain continuum free data cubes around a list of selected lines. 
 
We only focus here on molecular emission that originates  
from scales typically smaller than the largest angular scales of 
 $\sim$\,7\arcsec, beyond which  the sensitivity of our 12\,m array observations drops. We have therefore not used information from the more {compact 7m array}. We measure a 1$\sigma$ rms noise level of $\sim$3\,K in brightness temperature ($T_{\rm b}$) scales in a spectral resolution of 0.977\,MHz.
The spectral setup and data reduction for our target is described in more details in \citetalias{Csengeri2018}.

%-------------------------------------- Two column figure (place early!)

\begin{table*}[!ht]
\centering
\caption{List of positions, their observed 345\,GHz continuum emission and modelled methanol parameters.  }\label{tab:params}
\begin{tabular}{rcccccc}
\hline\hline
Position  & Offset$^{a}$ & $F_{\rm \nu}$$^b$ & $T_{\rm cont}$$^{c}$ & $N$(\methanol)$^\dagger$ & $T_{\rm kin}$$^{d,\dagger}$ & $N_{\rm H_2}$\\
		& 		& [mJy/beam] 		 & [K] & [cm$^{-2}$] & [K] & [cm$^{-2}$] \\
\hline
Shock \emph{A} & [$-1\rlap{.}{\arcsec}96$;$-2\rlap{.}{\arcsec}85$] & 156 & 30.9 & $1.2\times10^{19}-{\bf 1.6\times10^{19}}$ & 170-190 ({180}) & $2.04\times10^{24}$\\
Shock \emph{B} & [$-2\rlap{.}{\arcsec}50$;$-2\rlap{.}{\arcsec}88$] & 144 & 28.5 & $1.6\times10^{19}-2.4\times10^{19}$ (${\bf 2\times10^{19}}$) & 170-200 ({180}) & $1.89\times10^{24}$\\
Inner envelope   & [$-2\rlap{.}{\arcsec}43$;$-2\rlap{.}{\arcsec}47$] & ~~86 & 17.0 & ${\bf 4.1\times10^{18}}-8.1\times10^{18}$ & 90--{110} & $1.90\times10^{24}$ \\
Protostar + disk   & [$-2\rlap{.}{\arcsec}17$;$-2\rlap{.}{\arcsec}76$] & 239 &  47.3 & & $200-500$ & $1.1-2.8\times10^{24}$$^{e}$ \\

\hline
\end{tabular}\\
\tablefoot{\\
\tablefoottext{$\dagger$}{Bold face shows the values from the best fit models.}\\
\tablefoottext{a} {The position of the dust continuum peak is at [$-2\rlap{.}{\arcsec}17$;$-2\rlap{.}{\arcsec}76$] relative to the phase centre position given in Sec.\,\ref{sec:obs}.} \\
\tablefoottext{b} {Measured in the synthesized beam of $0\rlap{.}{\arcsec}3$$\times$$0\rlap{.}{\arcsec}17$.} \\
\tablefoottext{c} {Brightness temperature of the continuum emission on the Rayleigh-Jeans scale.} \\
\tablefoottext{d} {$T_{\rm kin}$ corresponds to the gas temperature estimated from the \methanol\ lines (Sect.\,\ref{sec:method}).} \\
\tablefoottext{e} {Neglecting the effect of optical depth.} \\
}
\end{table*}

\begin{table*}[!ht]
\caption{Summary of the detected interstellar COMs towards the selected positions.}\label{tab:lines2}
\centering
\begin{tabular}{lcccccrrrrrrrr}
\hline\hline
\multicolumn{2}{c}{Molecule} & $E_{\rm up}/k^d$& Detected $E_{\rm up}/k$ & Database & Component$^e$ \\

& & (K)& (K) &  &  \\

\hline
%CH$_3$OH, ${\varv_{\rm t}}$=0,1,2	   & methanol & 45 - 488 & JPL & e,s \\ 
CH$_3$OH	   & methanol & 45 -- 1677 & 45 -- 488 & JPL & e,s \\ 
CH$_3$OCHO& methyl formate &  80 -- 352 & 80 -- 352 & JPL & e,s \\
C$_2$H$_5$OH      & ethanol 	& 88 -- 3294 & 90  -- 407& CDMS & s \\
CH$_3$COCH$_3$$^a$     & acetone &  98 -- 2241 & 98 -- 304 &JPL & s \\ 
a(CH$_2$OH)$_2$		& ethylene glycol & 98 --1729 & 266 -- 455& CDMS & s \\
CH$_3$CHO     &  acetaldehyde & 153 -- 807 & 153 -- 383 & JPL & s \\
CH$_3$NCO     &  methyl isocyanate & 323 -- 465 & 365 - 460 & CDMS & s \\
C$_2$H$_3$CN & vinyl cyanide &  300 -- 2531 & 308 -- 326 & CDMS & c \\ 
C$_2$H$_5$CN	& ethyl cyanide & 60 -- 2032 & 72 -- 351 & CDMS & c \\  % done
HC(O)NH$_2$$^b$ & formamide  &  136 -- 152 & 136 -- 152 & CDMS & e,s,c \\ 
\hline
{\it{t}}-HCOOH$^{b,c}$		& formic acid &  136 -- 645 &145 & CDMS &  e   \\
\hline
\end{tabular}\\
\tablefoot{\tablefoottext{a} {Spectroscopic parameters are more uncertain at this frequency range \citep{Ordu2019}.} \\
\tablefoottext{b} {The identification is based on only one to two unblended transitions.}\\
{\tablefoottext{c} {Formic acid contains less than 6 atoms, it is, however, the simplest organic acid and thus will be discussed together with the listed interstellar COMs.}}\\
\tablefoottext{d} {Range of upper energy levels of all transitions in our frequency coverage with an $A_{\rm ij}$ above $10^{-4}$\,s$^{-1}$, except for methanol and methyl formate, where the lowest energy transition detected has an $A_{\rm ij}$ of $9.87\times10^{-5}$\,s$^{-1}$ and $2.68\times10^{-5}$\,s$^{-1}$, respectively.}\\
\tablefoottext{e} {The label \emph{e} corresponds to the inner envelope, \emph{s} to the accretion shocks, and \emph{c} to the innermost central regions that could have contribution  from both the outflow and the accretion disk. }

}
\end{table*}

%-------------------------------------- Two column figure (place early!)
  
\begin{table*}[!ht]
\caption{Summary of the model results for the detected COMs.}\label{tab:results}
\centering
\begin{tabular}{llcccccrrrrrrrr}
\hline\hline
Position & {Molecule} & $C_{\rm vib}$$^a$ & $N$/$N$(\methanol) & $N$ & $X^b$& $T_{\rm rot}$ & $v_{\rm off}$$^c$  \\
& &  & & [cm$^{-2}$] & & [K] &  [km\,s$^{-1}$]\\
\hline
Shock - A & CH$_3$OH	 & 1 & 1 & $1.6\times10^{19}$ &$7.8\times10^{-6}$  & 180 & $-4.2$ \\ %-> modified to the correct one, in the script, this is -jpl
		& CH$_3$OCHO&  1 & $0.04$&  $6.4\times10^{17}$ & $3.1\times10^{-7}$ & 180 & $-4.2$ \\
		& C$_2$H$_5$OH       & 1.44 & $0.048$ & $7.7\times10^{17}$ & $3.8\times10^{-7}$& 180 & $-4.2$ \\
		& CH$_3$COCH$_3$     &  1 & $0.025$ & $4.0\times10^{17}$  & $2.0\times10^{-7}$ & 180& $-4.2$  \\ 
		& a(CH$_2$OH)$_2$		&  1 & $0.005$  & $8.0\times10^{17}$  & $3.9\times10^{-8}$ & 180 & $-4.2$ \\
		& {\it{t}}-HCOOH$^b$		&  1 & $0.004$  & $6.4\times10^{16}$ & $3.1\times10^{-8}$ & 180 & $-4.2$ \\
		& CH$_3$CHO     &   1 & $0.002$  & $3.2\times10^{16}$ & $1.6\times10^{-8}$  & 180 & $-4.2$  \\
		& CH$_3$NCO     &   1.03 & $0.0041$  & $6.6\times10^{16}$ &   $3.2\times10^{-8}$ & 180 & $-4.2$ \\
		& C$_2$H$_3$CN &   1 & $0.0017$  & $2.7\times10^{16}$ & $1.3\times10^{-8}$ & 180 & $-4.2$  \\ 
		& C$_2$H$_5$CN	&  1.7 & $0.0057$   & $9.1\times10^{16}$ & $4.4\times10^{-8}$&  180  & $-4.2$ \\  % done
		& HC(O)NH$_2$$^b$ &  1.16 & $0.0058$ & $9.3\times10^{16}$ & $4.6\times10^{-8}$ & 180  & $-4.2$ \\ %-> has only one bright transition
		& CH$_3$OCH$_3$ &  1 & $<0.1$ & $<1.6\times10^{18}$ & $<7.8\times10^{-7}$ & 180 & $-4.2$ \\ %-> has only one bright transition
		& HDO &  1 & $0.02$ & $3.2\times10^{17}$ & $1.6\times10^{-7}$ & 180 & $-4.2$ \\ %-> has only one bright transition
 \hline
 
 Shock - B & CH$_3$OH	 & 1 & 1 & $2.0\times10^{19}$ &  $1.1\times10^{-5}$  & 180   & +4.55 \\ 
		& CH$_3$OCHO&  1 & $0.05$&  $1.0\times10^{18}$ & $5.3\times10^{-7}$ & 180 & +4.55\\
		& C$_2$H$_5$OH       & 1.44 & $0.048$ & $9.6\times10^{17}$ & $5.1\times10^{-7}$& 180 & +4.55 \\
		& CH$_3$COCH$_3$     &  1 & $0.025$ & $5.0\times10^{17}$  & $2.6\times10^{-7}$ & 180 & +4.55\\ 
		& a(CH$_2$OH)$_2$		&  1 & $0.005$  & $1.0\times10^{17}$  & $5.3\times10^{-8}$ & 180& +4.55 \\
		& {\it{t}}-HCOOH$^b$		&  1 & $0.005$  & $1.0\times10^{17}$  & $5.3\times10^{-8}$ & 180  & +4.55\\
		& CH$_3$CHO     &   1 & $0.002$  & $4.0\times10^{15}$ & $2.1\times10^{-9}$  & 180 & +4.55 \\
		& CH$_3$NCO     &   1.03 & $0.0041$  & $8.2\times10^{16}$ &   $4.4\times10^{-8}$ & 180 & +4.55\\
		& C$_2$H$_3$CN &   1 & $0.0017$  & $3.3\times10^{16}$ & $1.8\times10^{-8}$ & 180& +4.55 \\ 
		& C$_2$H$_5$CN	&  1.7 & $0.0049$   & $9.7\times10^{16}$ & $5.1\times10^{-8}$&  180& +4.55 \\  % done
		& HC(O)NH$_2$$^b$ &  1.16 & $0.0077$ & $1.6\times10^{17}$ & $8.2\times10^{-8}$ & 180 & +4.55\\ %-> has only one bright transition
		& CH$_3$OCH$_3$ &  1 & $<0.1$ & $<2.0\times10^{18}$ & $<1.1\times10^{-6}$ & 180 & $+4.55$ \\ %-> has only one bright transition
		& HDO &  1 & $0.02$ & $4.0\times10^{17}$ & $2.1\times10^{-7}$ & 180 & +4.55\\ %-> has only one bright transition
 \hline
 %%
 % H2 Cdensity at 110 K and 0.086 Jy/beam is 1.9d24 pp /cm^-2.
%
Envelope & CH$_3$OH	 & 1 & 1 & $ 4.1\times10^{18}$ & $2.2\times10^{-6}$  & 110 & $+3.5$ \\ %-> modified to the correct one, in the script, this is -jpl
		& CH$_3$OCHO&  1 & $0.025	$&  $1.03\times10^{17}$ & $5.4\times10^{-8}$  & 110  & $+3.5$ \\
		& C$_2$H$_5$OH       & 1.1 & $0.022$ & $9.0\times10^{16}$ &  $4.7\times10^{-8}$ & 110 & $+3.5$  \\
		& CH$_3$COCH$_3$     &  1 & $0.01$ & $4.1\times10^{16}$  & $2.2\times10^{-8}$ & 110 & $+3.5$ \\
		& a(CH$_2$OH)$_2$		&  1 & $0.01$  & $4.1\times10^{16}$  &  $2.2\times10^{-8}$ & 110 & $+3.5$ \\
		& {\it{t}}-HCOOH$^b$		&  1 & $0.0033$  & $1.367\times10^{16}$ & $7.2\times10^{-9}$  & 110 & $+3.5$ \\
		& CH$_3$CHO     &   1 & $0.00167$  & $6.83\times10^{15}$ &  $3.6\times10^{-9}$  & 110& $+3.5$   \\
		& CH$_3$NCO     &   1.0 & $0.004$  & $1.64\times10^{16}$ &   $8.6\times10^{-9}$ & 110 & $+3.5$ \\
		& C$_2$H$_3$CN &   1 & $0.005$  & $2.05\times10^{16}$ &   $1.1\times10^{-8}$& 110 & $+3.5$ \\ 
		& C$_2$H$_5$CN	&  1.18 & $0.0118$   & $4.8\times10^{16}$ & $2.6\times10^{-8}$ &  110 & $+3.5$  \\  % done
		& HC(O)NH$_2$$^b$ &  1.03 & $0.01$ & $4.2\times10^{16}$ &  $2.2\times10^{-8}$ & 110  & $+3.5$ \\ %-> has only one bright transition
		& CH$_3$OCH$_3$ &  1 & $<0.5$ & $<2.1\times10^{18}$ & $<1.1\times10^{-6}$ & 110 &  $+3.5$ \\ %-> has only one bright transition
		& HDO &  1 & $0.05$ & $2.1\times10^{17}$ &  $1.1\times10^{-7}$  & 110 & $+3.5$ \\ %-> has only one bright transition
\hline

\hline
\end{tabular}
\tablefoot{\\
\tablefoottext{a} {Correction factor to the molecular column density to account for the contribution of torsionally or vibrationally excited states to the partition function. All parameters are calculated for the given $T_{\rm kin}$.\\}
\tablefoottext{b} {The molecular abundance ($X$) is calculated by $X$=$N$/$N$(H$_2$). The values used for $N$(H$_2$) are given in Table\,\ref{tab:params}.\\}
\tablefoottext{c} {Velocity offset compared to the $\varv_{\rm lsr}$ of $-43.5$\,\kms.}\\
}
\end{table*}

   \begin{figure*}
   \centering
            {\includegraphics[width=0.95\linewidth]{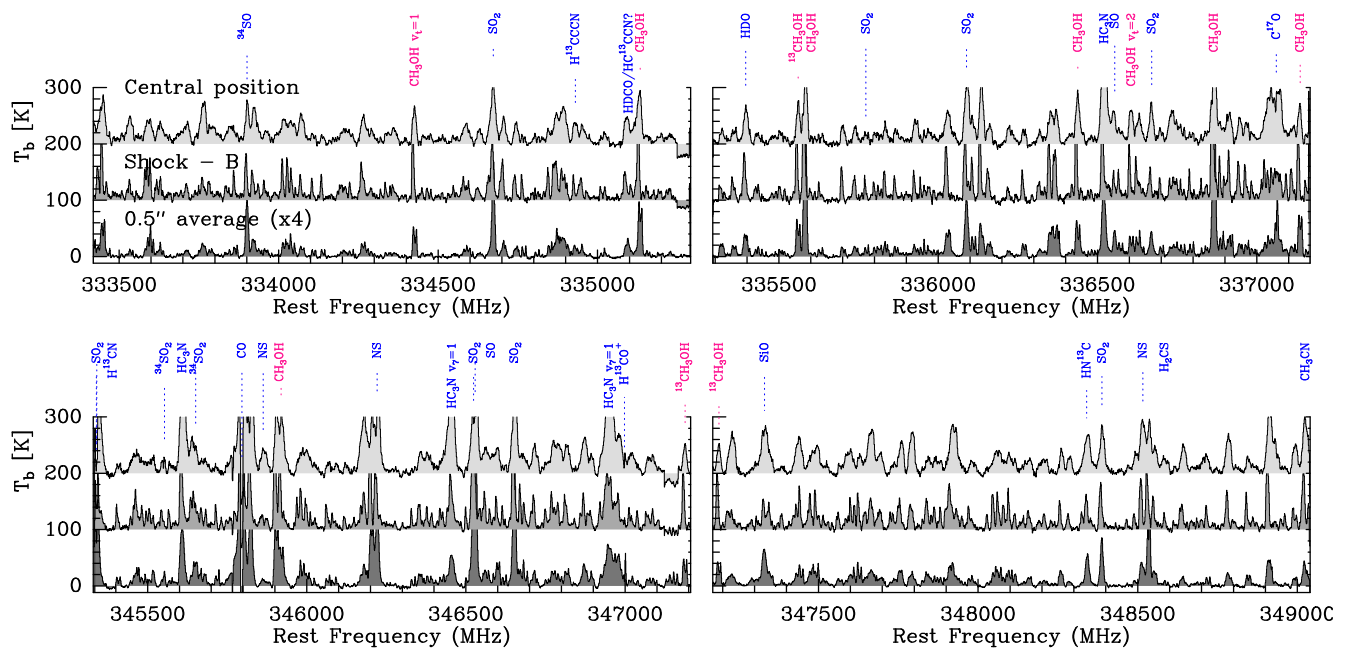}}
      \caption{{Grey filled histograms show spectra converted to brightness temperature ($T_{\rm b}$) scale: spectrum extracted towards the central position, the \textsl{B}-shock position, and an average within a radius of $0\rlap{.}{\arcsec}5$ around the peak of the dust continuum emission, which is multiplied by a factor of 4 for a better visibility}. The blue labels show the species listed in Table\,\ref{tab:lines}, pink labels show transitions of \methanol\ (and its isotopologue) discussed in \citetalias{Csengeri2018}. The spectra have been shifted along the $y$-axis for a better visibility.}
         \label{fig:average}
   \end{figure*}
  
%
%-------------------------------------------------------------
%                                             Two column Figure 
%-------------------------------------------------------------
   %\begin{figure*}
   \begin{sidewaysfigure*}
   \centering
            {\includegraphics[width=0.48\linewidth]{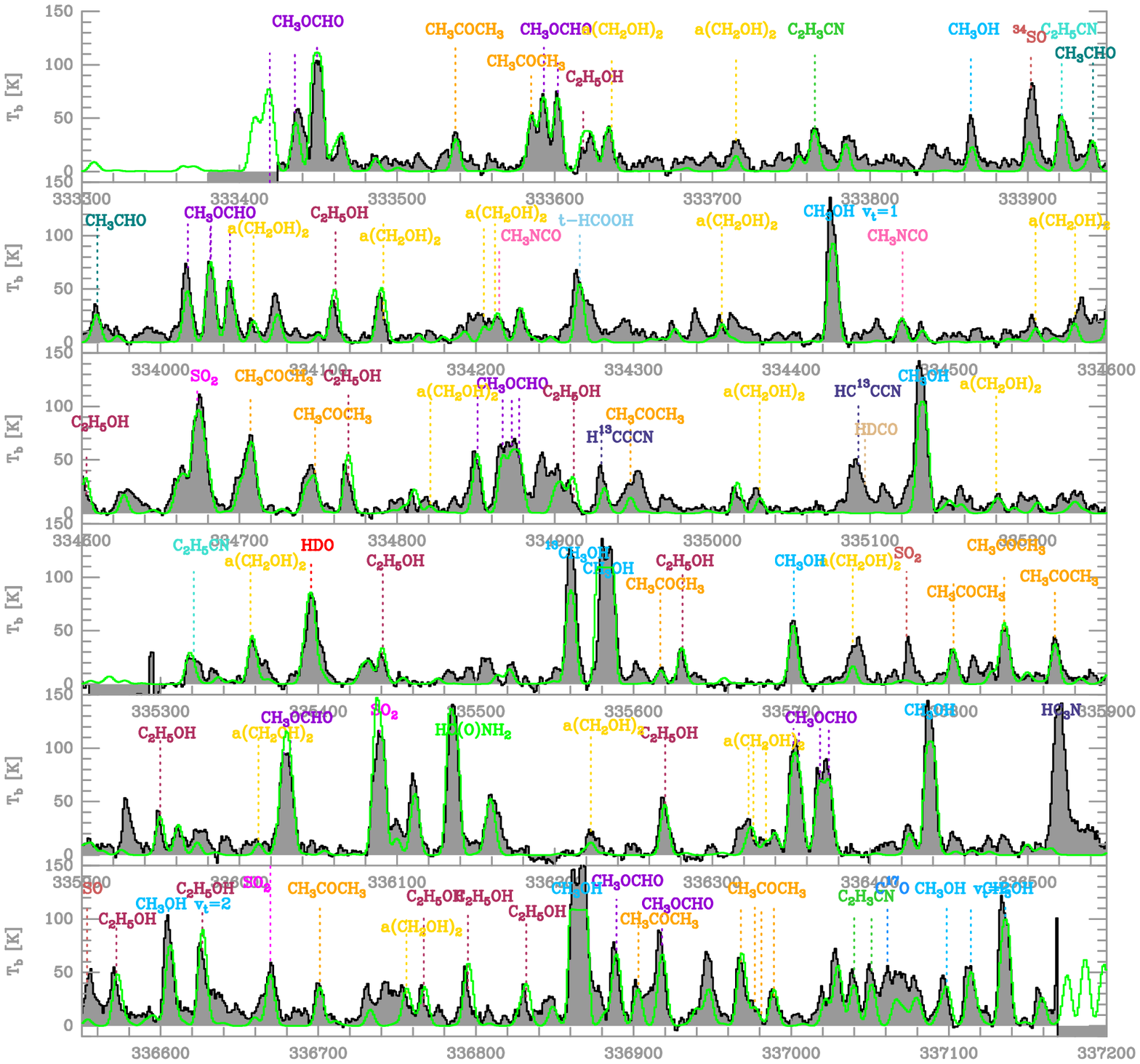}}
      %\resizebox{\hsize}{!}
         {\includegraphics[width=0.48\linewidth]{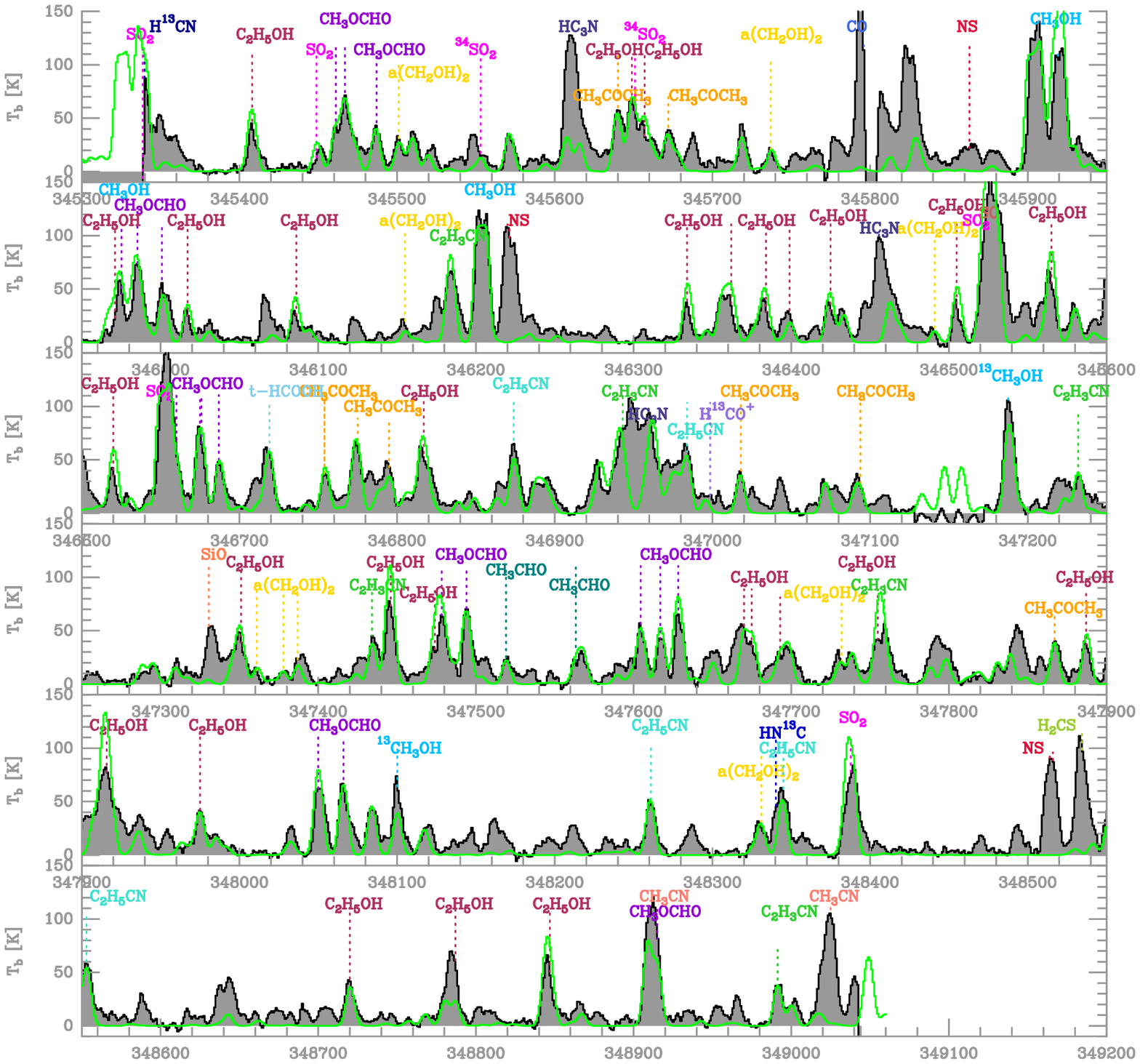}}
      \caption{Spectrum extracted towards the \emph{B}-shock position from \citetalias{Csengeri2018}. Grey filled histogram shows the observed spectrum converted to brightness temperature ($T_{\rm b}$) scale, green line shows the composite LTE model including the species listed in Table\,\ref{tab:lines2}. Coloured labels indicate the least blended transitions of the modelled COMs, as well as other species identified in the spectrum (Table\,\ref{tab:lines}). Each colour represents a different molecule and includes its isotopologues as well.}
         \label{fig:spectra}
  % \end{figure*}
   \end{sidewaysfigure*}
%
%-------------------------------------------------------------

\section{Results and analysis}\label{sec:results}

{We investigate here the physical properties of the gas and aim to explore its molecular composition with a particular focus on COMs within this high-mass protostellar envelope}. {In Fig.\,\ref{fig:average}, we show the spectrum towards the central position averaged within a radius of $0\rlap{.}{\arcsec}5$ (Fig.\,\ref{fig:overview}), and as a comparison the spectrum extracted towards the {accretion shock position labelled as} \emph{B}, and the central position corresponding to the protostar and its accretion disk. It is clear that the gas is rich in molecular emission, and we identify and list the brightest transitions of simple molecules in Table\,\ref{tab:lines}. }

%Here we aim to study and compare the molecular composition of the gas . 
We present a quantitative analysis towards three positions corresponding to the two locations of accretion shocks (one of them shown in Fig.\,\ref{fig:average}), and the bulk of the inner envelope  represented by a position offset from both the accretion shocks and the outflow impacted gas (Fig.\,\ref{fig:overview}). The severe blending due to the larger line-widths and the significant uncertainty of dust temperature and opacity hinders us from a quantitative analysis of the spectrum towards the central position, nevertheless, we show and qualitatively discuss its molecular emission corresponding to the protostar and its accretion disk.  We show the extracted spectrum towards the \emph{B}-shock position with the brightest emission lines, and hence the richest spectrum {in detail in Fig.\,\ref{fig:spectra} covering a 7.5\,GHz bandwidth where we also label  unblended transitions of the identified COMs}. The same figure for the other positions, corresponding to the \emph{A}-shock position and the position representing the bulk emission of the inner envelope, are shown in Appendix \ref{app:other_positions}.  %Using LTE modeling, we could identify the majority of the observed spectral lines  (Sect.\,\ref{sec:method}), allowing us to detect emission from complex organic molecules (Sect.\,\ref{sec:images}). Together with the line identification we also discuss their spatial distribution (Sect.\,\ref{sec:maps}). The modelling allowed us to identify the $J$=$3_{3,1}-4_{2,2}$ transition from HDO at 335.3955\,GHz (Sect.\,\ref{sec:hdo}).
 
\subsection{Line fitting and LTE modelling}\label{sec:method}
Since the volume density in the inner envelope is expected to be high ($\bar{n}\gtrsim10^7$\,cm$^{-3}$, \citetalias{Csengeri2018}), we model the spectra assuming that local thermodynamic equilibrium (LTE) conditions apply. We used the Weeds package \citep{Maret2011}, and fitted the spectra in an iterative process. Corrections for the torsionally and vibrationally excited states' contribution to the rotational partition function, and thus to the column density, were derived for the main isotopologue of the listed species. 

\begin{figure*}[!ht]
%   \resizebox{\hsize}{!}
            {\includegraphics[width=0.9\linewidth]{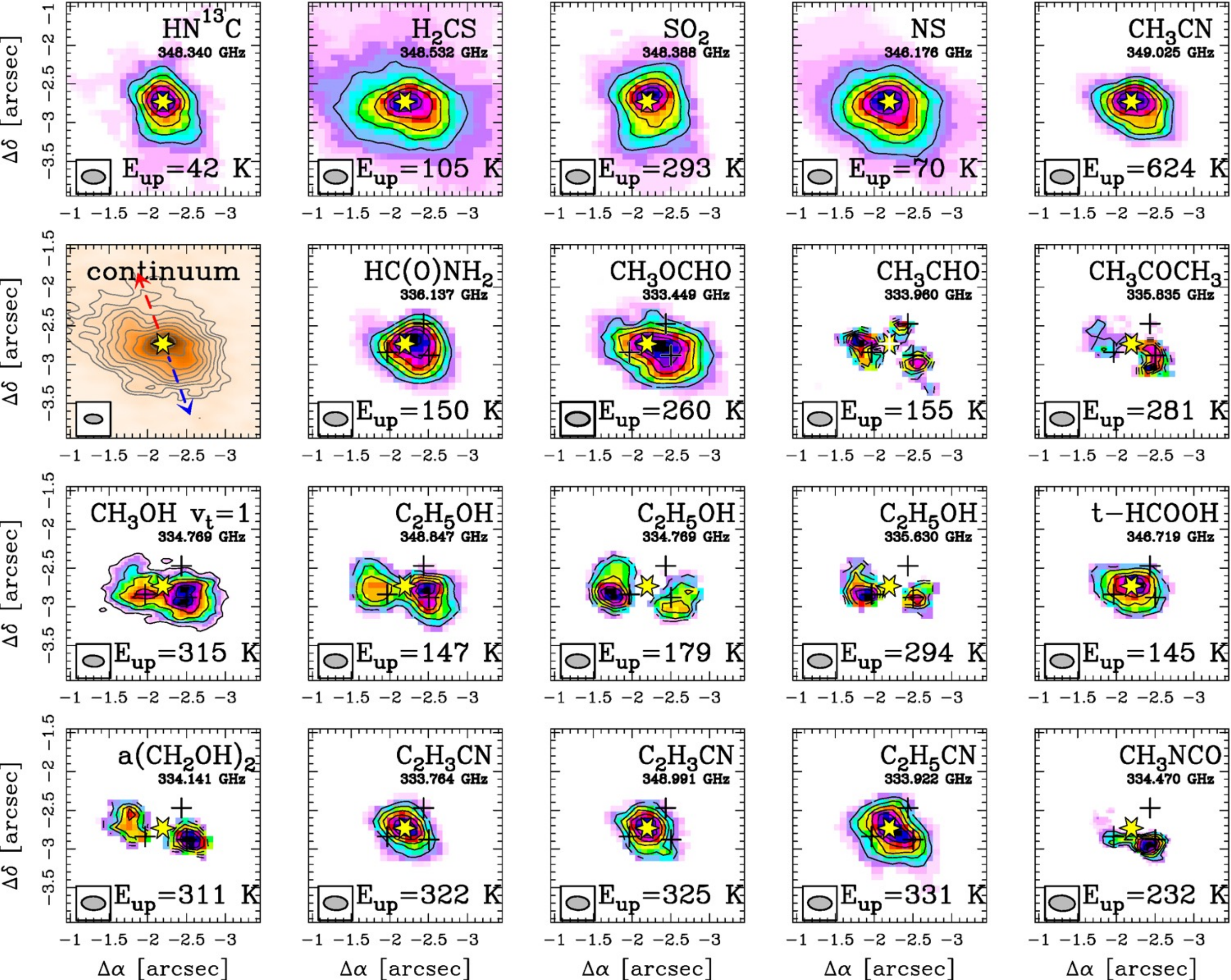}}
      \caption{Zeroth moment maps of some simple molecules, and selected transitions of COMs. {{All contours correspond to significant emission and}} start at {{20\%}} of the maximum value and increase by 15\% of the maximum value. Yellow star marks the position of the dust continuum peak corresponding to the central protostar. The beam is shown in the lower left corner. Labels indicate the shown molecular species. One figure shows the continuum as a reference, where the scale and contours are the same as in Fig.\,\ref{fig:overview}.}
         \label{fig:molecules_all}
   \end{figure*}

The steps of the fitting are the following. We first determined the molecular column density ($N$) and the kinetic temperature ($T_{\rm kin}$) of \methanol. For this the input parameters are $N$(\methanol), $T_{\rm kin}$, source size, rest velocity ($\varv_{\rm lsr}$) and linewidth ($\Delta \varv$).  {We fixed the source size to $0\rlap{.}{\arcsec}4$ which is larger than the beam.} While emission for  the  modelled transitions may have
different source sizes, as long as they are spatially resolved, the actual source size does not {significantly} influence the result. Later, from Fig.\,\ref{fig:molecules_all}, it is clear that this assumption holds, although emission from the \emph{B}-shock position seems to be more compact compared to the other positions. Overestimating the source size would lead to an underestimation of the molecular column densities.
The $\varv_{\rm lsr}$  and $\Delta \varv$ are measured using a Gaussian fit to the CH$_3$OH lines and these values are reported in \citetalias{Csengeri2018}.  We also considered the continuum emission in our models that we have directly extracted from the fitted baseline of the spectra. These parameters, together with the position, the measured continuum flux density, and the results of the two free parameters, $N$(\methanol) and $T_{\rm kin}$ are summarised in Table\,\ref{tab:params}.

To constrain $N$(\methanol) and $T_{\rm kin}$, we created a grid of models exploring a parameter range of $N$(\methanol)=$10^{17}-10^{20}$\,cm$^{-2}$ and $T_{\rm kin}$=50$-$300\,K. To select the best fit, we computed the residuals of each fit, and visually inspected the results with a particular emphasis on the optically thin lines and the unblended transitions of \methanol\ {listed in Table\,\ref{tab:lines}}. The resulting $T_{\rm kin}$ was then used to estimate $N_{\rm H_2}$ based on the dust continuum emission assuming that the gas and the dust are thermalised at these high densities, hence $T_{\rm d}=T_{\rm kin}$. For this we used $N$(H$_2$) = $\frac{F_\nu\,R}{B_\nu(T_d)\,\Omega\,\kappa_{\nu}\,\mu_{\rm H_2}\,m_{\rm H}}
\rm{[cm^{-2}]}$, where $F_\nu$ is the beam averaged flux density towards the selected positions, $B_{\rm \nu} (T)$ is the Planck function, 
$\Omega$ is the solid angle
of the beam calculated by $\Omega = 1.13\times \Theta^2$, where 
$\Theta$ is the geometric mean of the beam major and minor axes; 
$\kappa_{\nu}=0.0185$\,cm$^{2}$\,g$^{-1}$ from \citet{OH1994} {at 345 GHz} including the gas-to-dust ratio, R, of 100;
 $\mu_{\rm H_2}$ is 
 the mean molecular weight per 
hydrogen molecule and is equal to 2.8; and
 $m_{\rm H}$ is the mass of a hydrogen atom.

%We find a good fit to the data\footnote{Since the dust emission is likely to be optically thin at the positions off the dust continuum peak, excluding the continuum level would only marginally affect our results, giving a parameter range of 140-170\,K and $N$(\methanol)=$1.2-2.0\times10^{19}$\,cm\,$^{-2}$.} within a parameter range of $T_{\rm kin}$=$160-200$\,K, and $N$(\methanol)=$1.6-2.4\times10^{19}$\,cm\,$^{-2}$ on the shock position marked as \emph{A}. Using the same approach, on the other position towards the inner envelope (offset from the protostar, and the shock position), we find $T_{\rm kin}$=$90-130$\,K and $N$(\methanol)=$4.1\times10^{18}$\,cm\,$^{-2}$.

We find that the derived parameters are similar for the two shock positions, the \methanol\ column density reaches up to $2\times10^{19}$\,cm$^{-2}$, and the kinetic temperature is around 180\,K. As pointed out in \citetalias{Csengeri2018}, temperatures above $\sim$200\,K are not consistent with the observations because other \methanol\ lines would appear brighter than the emission seen at the corresponding frequencies. Towards the position of the inner envelope, our models suggest a factor of 5 lower \methanol\ column density, and a considerably lower kinetic temperature of 110\,K. In Table\,\ref{tab:params} we give the range of values {that are consistent with the data within 3$\sigma$}, and mark in bold the {best fit} value adopted for the following modelling.

After {fitting} the \methanol\ transitions, we created the models for other molecules using the same parameters (\vlsr, $\Delta \varv$, $T_{\rm kin}$ and source size) as for \methanol. Varying the molecular column density (relative to methanol) we fitted the brightest unblended transitions of each species and visually inspected the result. 
As shown later in  Fig.\,\ref{fig:molecules_all}, the adopted source size corresponding to spatially resolved emission holds  for most of the molecules, however, some of the highest energy transitions seem to originate from a more compact region which we do not account for in our model. As a consequence, the column densities for these molecules may be underestimated.  {As our composite final model shows, these parameters give a good agreement with the spectra. The line-width of vinyl and ethyl cyanide are, however, larger than that of methanol with line-widths up to 8\,\kms. Such a larger value would lead to molecular column densities up to 50\% higher than the value given by our model.}

%Therefore we add the other molecular components with the same physical parameters  and only vary their relative abundance compared to \methanol. 
%The column density of each specie was adjusted to fit the unblended transitions of the spectra. 
We obtained the final model in an iterative process, by first fitting the brightest emission of a new species independently, and then visually inspected the full model including all identified species and checked whether the overall fit result remains reasonably good. This is necessary due to the large fraction of blended transitions. %This way we could also identify which transitions of the selected species are affected by blending. 
Subsequently, once a molecule was identified, we included its isotopologue in the model, and searched for rotational transitions of their lowest energy vibrationally excited states. However, except for \methanol\ $\varv_{\rm t}$ = 1 and HC$_3$N $\varv_7$=1, we have no clear detection of emission from such higher energy, {vibrationally excited} states. 

Altogether we could identify emission from 10 COMs in the spectra. We list the results of the fit, such as their relative fraction compared to \methanol, molecular column density, and abundance in Table\,\ref{tab:results} for the three positions. %Finally, to obtain a reasonably good fit to the full spectra, we included additional species such as HC$_3$N and S-bearing molecules, however, their fitted parameters are less physically motivated and we only aimed at a visually good fit to their lines.
%We used the same kinetic temperature assuming that all the observed species originate from the same gas, 
We show the spectrum for the \emph{B}-shock position and the best-fit model in Fig.\,\ref{fig:spectra}, where we label the least blended lines of each molecule. 

%-------------------------------------------------------------
%
%-------------------------------------------------------------

\subsection{Detection of COMs}\label{sec:images}

While the available, non continuous 7.5\,GHz bandwidth offers limited spectral coverage to perform a complete analysis of all the molecules that have lines in the spectra, we are able to identify several transitions from ten COMs {(Table\,\ref{tab:lines2}). 
This frequency range covers transitions over a broad range of upper energy levels ($E_{\rm up}/k$) with a spontaneous decay rate (Einstein A-coefficient) 
above $10^{-4}$\,s$^{-1}$ for all species. The range of upper energy levels is, however, much narrower for the detected transitions and covers a range typically around 150 to 400~K. Many of these transitions are, however strongly blended in the spectra.  The smallest energy range covered is for methyl formate and formamide.}% with the lowest column density (CH$_3$CHO) of 4$\times10^{15}$\,cm$^{-2}$ towards the \emph{B}-shock position. 

In addition to several methanol (and isotopologue) lines, we detect emission from various O-bearing COMs, such as methyl formate (CH$_3$OCHO $\varv$=0), ethanol (C$_2$H$_5$OH $\varv$=0), acetone (CH$_3$COCH$_3$ $\varv$=0), ethylene glycol ($a$(CH$_2$OH)$_2$), and acetaldehyde (CH$_3$CHO,$\varv$=0). {Comparing this list of O-bearing COMs with the typical molecules identified towards hot cores, a notable non detection is dimethyl ether (CH$_3$OCH$_3$) which is often found to be co-existing and spatially correlated with methyl formate (e.g.\,\citealp{Brouillet2013, Jaber2014}) suggesting that they are chemically related \citep{Garrod2008}. The upper limit on its molecular column density suggests, however, that a significant amount of dimethyl ether could remain undetected in this frequency range (see Sect.\,\ref{sec:d1}).}

Among the detected species except methanol, methyl formate has the largest column density up to $N$(CH$_3$OCHO)=1.0$\times10^{18}$\,cm$^{-2}$ corresponding to a relative fraction of $0.05$ compared to \methanol. The largest number of transitions falling in this band are from acetone, ethylene glycol and acetaldehyde which produce a weed-like spread of spectral lines. Therefore, due to blending and uncertainties in the spectroscopic parameters at such relatively high energy and frequency \citep{Ordu2019}, the identification of these molecules is less robust. %For example, as shown in Fig.\,\ref{fig:acetone_lines}, we find the largest difference between the modelled and the observed spectra for acetone. 
As a consequence, and also considering their large fraction of blended transitions, the estimated column density is less robust compared to the other species. In particular, blending seems to be more prevalent for   acetone, ethanol and acetaldehyde transitions in the higher frequency spectral windows between 345 and 349\,GHz. %In App.\,\ref{app:lists} we list the transitions of each molecule with $E_{\rm up}<500$\,K\footnote{We find that in our model no transition with $E_{\rm up}>500$\,K would be detectable for COMs.}, and show the observed and modelled spectra of the corresponding frequency range.

We also likely detect formic acid, $t-$HCOOH, however, there are only two detectable lines of this molecule falling in our frequency coverage and only one of them is unblended\footnote{The  $^{13}$C isotopologue of formic acid has an expected peak line temperature comparable to the noise level.}. {Our modelling suggests that $t-$HCOOH is among the lowest abundance O-bearing COMs, with typically more than two orders of magnitude lower abundance compared to that of \methanol.}

%Due to blending, the relative abundance of acetone is more uncertain than the previous two species, and can be as high as 1/40 with respect to \methanol.

We also identify N-bearing COMs, such as vinyl cyanide (C$_2$H$_3$CN), and ethyl cyanide, (C$_2$H$_5$CN), methyl isocyanate (CH$_3$NCO) as well as  formamide (HC(O)NH$_2$). The identification of formamide is, however, less robust since it has practically only one detectable transition that is not blended  with other lines in the observed frequency range. Since the unblended formamide line is strong, in order to increase the number of its detectable transitions, we searched for its $^{13}$C isotopologue both in the spectra extracted towards the shock spots and the brightest central position. The brightest isotopologue line ($J$=16$_{2,15}$--15$_{2,14}$) is, however, blended at the frequency of 335.405\,GHz, and has a predicted peak line temperature close to the noise level assuming an isotopic ratio of $60$ and optically thin emission. %It was therefore not possible to render our detection more robust, nevertheless, 
 
Comparing the molecular abundances between O- and N-bearing COMs, we find more than one order of magnitude lower column densities for the N-bearing COMs, such as vinyl  and ethyl cyanide, as well as formamide compared to the O-bearing COMs. {O-bearing COMs originating from a colder gas component have been seen on similar scales towards the Orion Hot Core and the Compact Ridge \citep{Caselli1993}, and on somewhat larger scales towards classical hot cores \citep{Qin2010,Weaver2017}.} However, our analysis in Sect.\,\ref{sec:maps} suggests that our initial assumption that all COMs originate from gas with the same physical conditions (i.e. $T_{\rm kin}$), may not hold,
{in particular for the central position where COMs with a CN group seem to be more compact with a different spatial origin.} In this case our estimation of the column density, especially for C$_2$H$_3$CN and C$_2$H$_5$CN may not be {accurate}.
%Other N- and S-bearing species, such as HC$_3$N, HN$^{13}$C, SO, and SO$_2$ are also detected, however, either they only have one transition falling in the band, or their distribution and kinematics suggest a different physical origin associated with the outflow of the source that is typical for example for the S-bearing species. Therefore we do not make an attempt here to include them in our model. % consistent with a purely rotating envelope/disk, except for the highest energy $v_7=1$ transition of  HC$_3$N. 
 % We assign the 333865.0 Unidentified line from Schilke et al. 1997 to C2H5CN.

\subsection{Spatial distribution of various molecules}\label{sec:maps}

{Investigating the spatial distribution of the identified molecules, we find a strong chemical differentiation within this high-mass envelope.} 
%To investigate the spatial distribution of the identified molecules, 
We show 0th moment maps calculated over a velocity range of $-55$ to $-35$\,\kms\  for some unblended transitions of COMs in Fig.\,\ref{fig:molecules_all}, and for comparison, we also show some of the brightest emission from other molecules, such as HN$^{13}$C (J=4--3, $E_{\rm up}/k$=42\,K), H$_2$CS (J=$10-{1,9}-9_{1,8}$, $E_{\rm up}/k$=105\,K), SO$_2$ (J=$24_{2,22}-23_{3,21}$, $E_{\rm up}/k$=293\,K), NS (J=$8_{1,8,7}-7_{-1,7,6}$, J=$8_{1,8,8}-7_{-1,7,7}$, $E_{\rm up}/k$=70\,K), as well as   the $K$=8--8 line of the $J$=19--18 transition of CH$_3$CN ($E_{\rm up}/k$=624\,K). All these transitions {of simple molecules} show the brightest emission towards the peak position of the dust continuum, and show  considerably more extended emission towards the inner envelope than the majority of the COMs. Only the shown CH$_3$CN line appears relatively compact highlighting the potentially warmer regions in the immediate vicinity of the protostellar embryo.

The emission from COMs is rather compact, but typically spatially resolved. {Methyl formate shows the most extended morphology coinciding with the highest column density dust emission, however, formamide is also considerably extended.} Using a 2D Gaussian fit to the 0th moment map shown in Fig.\,\ref{fig:molecules_all}, we measure the beam convolved full-width at half-maximum (FWHM) extent of methyl formate to be $0\rlap{.}{\arcsec}95$$\times$$0\rlap{.}{\arcsec}58$, corresponding to a beam deconvolved geometric mean of 0\rlap{.}{\arcsec}70. Following the formulation  in \citetalias{Csengeri2018}, we estimate a radius ($R_{\rm 90\%}$) of 0\rlap{.}{\arcsec}58 corresponding to a size of 1450\,au. For formamide, we measure a beam convolved FWHM  of $0\rlap{.}{\arcsec}62$$\times$$0\rlap{.}{\arcsec}61$, corresponding to a beam deconvolved geometric mean of $0\rlap{.}{\arcsec}57$.
Similarly as above, this corresponds to a radius of $0\rlap{.}{\arcsec}47$ that is 1175\,au. %on physical scales. 

{Qualitatively comparing the emission from the central position towards that of the envelope,} it is apparent that the O- versus N-bearing COMs show a striking difference in their morphology; while O-bearing COMs peak offset from the central protostar, N-bearing COMs with a CN group are the brightest towards the {central position, thus the protostar and the disk}, and show an elongation in the direction of the outflow axis. {The most striking example of this dichotomy is seen in ethanol, acetone, acetaldehyde and ethylene glycol versus vinyl and ethyl cyanides. The similar upper energy levels {($E_{\rm up}/k$=150$-$300\, K)} of these transitions (Fig.\,\ref{fig:molecules_all}) suggests that the spatial morphology corresponds to genuine chemical differentiation rather than temperature gradients and excitation effects.} 
Similarly as above, using a 2D Gaussian fit, we measure a beam convolved FWHM of the most compact emission of the 348.991\,GHz C$_2$H$_3$CN line of $0\rlap{.}{\arcsec}63$$\times$$0\rlap{.}{\arcsec}51$, corresponding to a beam deconvolved beam size of $0\rlap{.}{\arcsec}55$$\times$$0\rlap{.}{\arcsec}47$. The most compact component of this emission corresponds to an $R_{90\%}$ radius of $\sim$900\,au.

{The other striking feature is that, all the other typical O-bearing COMs, such as} ethanol, acetone, ethylene glycol peak on and only show emission towards the shock positions. The lack of detection {of these molecules} towards the inner envelope may be, however, due {to observational limitation, because the unblended transitions in our frequency coverage are typically above $E_{\rm up}/k$$>$ 200\,K, hence a combination of sensitivity, blending and excitation effects may make} it difficult to detect these molecules in the relatively colder gas component with $T_{\rm kin}$$\lesssim$100\,K.

\subsection{Detection and distribution of HDO}\label{sec:hdo}

%  1 HDO           335395.500   0.026   335.3    7  2.61e-05          3 3 1 -- 4 2 2          weeds_c019002.cat
The LTE modelling allows us to identify the $J$=$3_{3,1}-4_{2,2}$ transition of HDO at 335.396\,GHz ($E_{\rm up}/k=335$\,K), and requires a high column density of $2.1-4.0\times10^{17}$\,cm$^{-2}$
to fit the observed line intensity.
%a(CH2OH)2            335396.713   0.005   316.7  603  8.53e-04       33 924 0 -- 32 923 1       weeds_c062503.cat
Our models suggest that this transition is blended with a line of ethylene glycol at 335.397\,GHz. In addition, shifted by a few \kms\ at 335.403\,GHz, there is also a methyl formate line which may also show a small contribution to the observed emission at the frequency of the HDO line. 
{However, there is no model that could reproduce the observed spectrum using only ethylene glycol and methyl formate.} Our best fit model towards all three position requires including a significant amount of HDO to reproduce the observed line intensity. 
{The spatial extent of the HDO line is comparable to that of methyl formate and formamide, while ethylene glycol for example has a considerably more compact morphology peaking on the accretion shocks. The modelled lines do not contribute to more than 15-21\% of the velocity integrated line intensity on the selected positions, therefore we assign the observed emission to the HDO line.}

The distribution of the HDO emission together with its velocity field is shown in Fig.\,\ref{fig:hdo}. It shows an increased intensity towards one of the shock positions, however, its emission is dominated by the inner envelope component showing an extended morphology. {The estimated HDO abundance is between $1.1\times10^{-7}$ and $2.1\times10^{-7}$ for all three positions, and is therefore similar towards the shock positions and the inner envelope.} Using a 2D Gaussian fit we measure its beam convolved FWHM to be $\sim$$0\rlap{.}{\arcsec}86$$\times$$0\rlap{.}{\arcsec}54$ with a deconvolved geometric mean of $0\rlap{.}{\arcsec}64$ corresponding to an $R_{90\%}$ radius of $0\rlap{.}{\arcsec}53$ that is $\sim$1325\, au. As seen in Fig.\,\ref{fig:molecules_all}, this shows that the emission from HDO (and thus likely of that of  H$_2$O in the gas phase) is comparable to that of the inner envelope, as well as  that of methanol and methyl formate. Also formamide shows a similar, although marginally smaller extent.
Similarly to \methanol\, and other O-bearing COMs, its peak is offset compared to the position of the protostar, suggesting that it has a higher abundance at the shock positions. This is confirmed by our modelling which measures the highest column density of HDO towards the \emph{B}-shock position. In Fig.\,\ref{fig:hdo} we show the first moment map revealing the kinematics of the HDO emitting gas. The velocity pattern of HDO traced by its 1st moment map
 is very similar to that of \methanol\ \citepalias{Csengeri2018}, 
and is consistent with rotational motions with a velocity gradient roughly perpendicular to the outflow. This confirms that the bulk of observed HDO emission originates from the inner envelope.

%18 CH3OCHO,vt=0          335402.759   0.100    94.9   62  3.83e-05       15 6 9 0 -- 14 510 0       weeds_c060103.cat
 
% There may also be a weak contribution by acetnoe, but thi sis really weak.
%... molecule blending with the HDO line, however, these results also clearly show that the observed line intensity is not consistent with an origin entirely of ...
%Ethylene glycol has a transition blending with the HDO line. Our models have, however, poor fit to the other ethylene glycol transitions suggesting that we overestimate the column density of this specie.

\section{Discussion}\label{sec:discussion}

\subsection{Shock chemistry instead of a radiatively heated envelope: precursor of a hot core}\label{sec:d1}

In \citetalias{Csengeri2018} we investigate the physical properties of the collapsing core and the protostellar envelope of \mysou, and estimate the current protostellar mass to be around 16\,\msol\ with an envelope mass of $M_{\rm env}$$\sim$$130$\,\msol. Therefore, the protostar is very likely to form an O4-O5 type star with a final stellar mass of $\sim50$\,\msol. While the bulk of the  \methanol\ emission is extended over the inner envelope, at the close vicinity of the protostar, between a projected distance of 300 and 800\,au, a rotational transition from the vibrationally excited state of \methanol\ at 334.436\,GHz is interpreted as tracing shocks due to accretion. 
{The accretion shocks imply the presence of an accretion disk, with a measured orientation perpendicular within 10$^\circ$ to the axis of the outflow. Since the accretion shocks on both the A and B positions are extended, we study here the molecular composition of positions that are the brightest in molecular emission along the extent of the accretion shocks (Fig.\,\ref{fig:overview}).}

Our results on the overall molecular composition of the gas suggest a similar molecular richness compared to other high-mass star forming regions hosting classical, {radiatively heated} hot cores (\citealp{Hatchell1998,Bisschop2007, Allen2017, Weaver2017}).
{The COMs typical of hot cores are detected towards \mysou, however, several species are found to peak at the proposed accretion shocks rather than the radiatively heated core towards the protostar and the accretion disk. }
%However, the estimated physical parameters in \mysou\ suggest that the hottest region with $T$$>$$200$\,K must be confined to a  compact inner region. % with a small radius of $\lesssim300$\,au. 
{Inferred from \methanol\ transitions, the bulk of the gas in the inner envelope is at a temperature of $T_{\rm kin}=110$\,K
with a relatively small extent, a radius of $\sim$1175\,au, and 1450\,au for the most extended molecules, such as formamide and methyl formate, respectively (Sect.\,\ref{sec:maps}). This extent is comparable to the highest column density dust continuum emission detected with a 1500\,au radius. Considering
a spherically symmetric centrally illuminated core, the expected dust temperature is between 83 and 157\,K at a radius of 1000\,au for the protostellar luminosity of 1.3$\times$$10^4$\,\lsol\,\citepalias{Csengeri2018} assuming radiative equilibrium (\citealt{GoldreichKwan1974,WolfireCassinelli1986, Wilner1995}). This is broadly consistent with the temperature estimate from the molecular line emission for the bulk of the inner envelope. However, instead of a gradual warming up of the gas due to the radiative heating of the protostar, we identify localised spots of heated gas towards the accretion shocks with $T_{\rm kin}=180$\,K. }

{Since the source hosts a deeply embedded protostar, a radiatively heated inner core is expected towards the central position. The most compact emission peaking on the protostar is traced by vinyl and ethyl cyanides and has a radius of 900\,au along its minor axis. Assuming that molecules with a CN group outline the largest potential extent of a radiatively heated inner core implies that this region must be very compact.} Massive protostars or YSOs observed towards well studied high-mass star forming regions exhibit a considerably more extended heated inner region. For example \citet{Ginsburg2017} resolve the region with high gas temperatures of $T>100$\,K out to 5000\,au towards the hot cores of the W51 Main star forming region, {and \citet{Bonfand2017} measure an extent of $\sim3000$\,au towards compact hot cores in the SgrB2 star forming region}. Other hot cores typically have an even larger extent of warm gas up to 0.1\,pc \citep{Hatchell1998,Kurtz2000, Cesaroni2005}. 
{In this context the current state of \mysou\ is likely to represent an earlier evolutionary stage compared to classical hot cores, and where localised spots of heated gas due to accretion shocks, hence a different heating mechanism leaves an observable imprint on the physical and chemical properties of the gas, {since the highest column densities of COMs originate from the accretion shocks. }} 

{In the case of \mysou, based on the radiative equilibrium model, a $\sim$15$\times$ higher protostellar luminosity would be required to heat the dust to $T>100$\,K at a 5000\,au radius.} Given the large expected final stellar mass of $M_{\rm \star}\sim50$\,\msol\ of {the protostar in \mysou}, it is very likely that the radiatively heated region is going to expand, turning this object into a classical radiatively heated hot core with a comparably large extent of heated gas as other known objects. %hot core because as the mass of the central object grows, its radiative feedback is expected to progressively heat up a larger amount of gas increasing the extent of the hot gas dominated region. 

{{
In addition to these physical arguments, the molecular composition could also be used to obtain an age estimate of the gas (e.g.\,\citealp{Viti2004,Caselli1993,Garrod2008,Allen2018}). Overall, we detect several molecules that are suggested to be  "first generation" COMs originating from grain surface chemistry, based on the observational study of \citet{Bisschop2007}, such as CH$_3$OH, C$_2$H$_5$OH, CH$_3$OCHO, HC(O)NH$_2$, CH$_3$CN, and C$_2$H$_5$CN, although the spatial distribution of these molecules is resolved and shows significant differences (see Sect.\,\ref{sec:d2}). Chemical age estimates for hot cores have been based on molecular richness \citep{Calcutt2014}, ratios of chemically related species such as C$_2$H$_3$CN and C$_2$H$_5$CN (e.g.\,\citealp{Fontani2007,Zeng2018,Allen2018}), and dedicated physico-chemical modelling of hot cores \citep{Bonfand2019}.
%since C$_2$H$_3$CN is mostly produced after C$_2$H$_5$CN is evaporated from the grain surfaces. 
{The observed ratio of C$_2$H$_3$CN and C$_2$H$_5$CN towards \mysou\ is consistent with a source that is still chemically young according to the chemical models by \citet{Caselli1993}, although \citet{Charnley1992} and \citet{Rodgers2001} suggest that additional C$_2$H$_3$CN may form in gas phase reactions compared to those models. In this context the overall chemical composition of \mysou\ resembles that of a young hot core with an age $<$10$^5$\,years, and is different from sources where chemistry is driven by low velocity shocks and a high cosmic ray flux \citep{Zeng2018}. Recent models by \citet{Allen2018}, for example, reproduce the observed abundances of cyanides using a short warm-up phase, however, including a higher cosmic ionisation rate towards one hot core in their sample was necessary. Therefore, it remains unclear how reliable such age estimates can be, missing gas-phase reactions for the production of cyanides as well as variations in physical conditions may impact age estimates.  }  %The key result of our study shows, however, that due to the spatial differentiation of these molecules the chemical evolution of these components of the protostellar envelope may be different. 
}
%Note that here we only discuss the warm and hot gas phase!

%Conclusion: CH3OH, C2H5OH, HCOOCH3 could potentially result from first generation chemistry taking place in the ices on grain surfaces, although they have higher rotation temperatures. Even C2H5VCN could be the result of first generation ice production.

%-------------------------------------------------------------
   \begin{figure}
   \centering
  % \resizebox{\hsize}{!}
            {\includegraphics[width=0.95\linewidth]{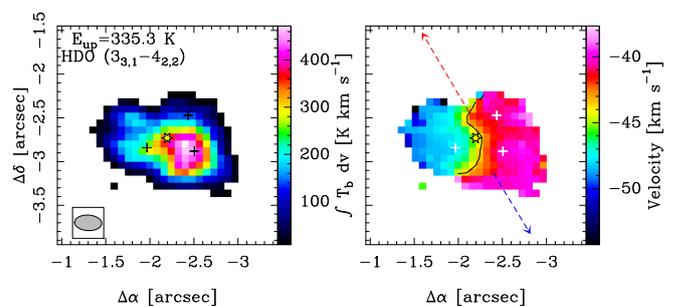}}
      \caption{Zeroth (left) and first (right) moment map of the HDO (J=$3_{3,1}-4_{2,2}$) line calculated between $-55$ and $-35$\,\kms. The yellow star marks the position of the dust continuum peak, and the black line shows the \vlsr\ at $-43.5$\,\kms. The red and blue arrows show the orientation of the outflow. The crosses mark the positions where the spectra were extracted for this study. The beam is shown in the lower left corner. }
         \label{fig:hdo}
   \end{figure}
%
%-------------------------------------------------------------

\subsection{Chemical differentiation {at} the innermost 1500\,au scales of the protostar: the O/N dichotomy}\label{sec:d1}

One of the most striking results of our analysis is the observed spatial dichotomy between O- and N-bearing COMs on scales smaller than $\sim$1000\,au observed towards the inner envelope. While the O-bearing COMs are associated with the inner envelope and the shock spots, the N-bearing COMs are located in the immediate vicinity of the protostar {and the accretion disk} and show an extension in the direction of the outflow. Only formamide shows a more extended distribution, similar to that of methyl formate.

Chemical differentiation, a.o., the O/N dichotomy, has been first observed on larger scales towards Orion-KL \citep{Caselli1993} as well as W3(OH) and W3(H$_2$O) \citep{Wyrowski1999} originating from different physical components within these star forming regions. {On scales smaller than 3000\,au this has been further confirmed towards Orion-KL \citep{Blake1996,Wright1996,WidicusWeaver2012,Friedel2012,Feng2015}, as well as other high-mass star forming cores, such as G10.61$-$0.23 \citep{Qin2010} followed by several other examples, like AFGL2591 \citep{Jimenez-Serra2012}, NGC7538IRS9 \citep{Oberg2013}, G35.20, G35.03 \citep{Allen2017}, and other sources like more evolved MYSOs \citep{Fayolle2015}. \mysou\ is, however, the first example where we can study such a chemical differentiation in a single, well resolved collapsing envelope, and associate the observed molecules to their physical origin within the envelope.}  
This makes this source a favourable target to study the origin of chemical differentiation as well as the emergence, and chemical evolution of hot cores.

% ABUNDANCES
From the O-bearing COMs except methanol, methyl formate has the largest column density, and its distribution resembles that of  the low-excitation \methanol\ lines shown in more detail in \citetalias{Csengeri2018}. This is, however, not surprising, since methyl formate is chemically related to methanol \citep{Garrod2006}.  The observed spatial distribution and its presence in a moderately warm gas phase may suggest grain surface production and subsequent sublimation to the gas phase. {Dimethyl ether, a chemically related molecule to methyl formate remains, however, undetected up to a relative fraction of 10-50\% compared to methanol towards both the shock and the envelope positions, despite having transitions in a range of upper energy levels similar to that of methyl formate. The models of \citet{Garrod2006} predict a ratio between the gas phase fractional abundances of methyl formate and dimethyl ether between 0.24 and 17 at different times of their models, while observations suggest a molecular abundance of the same order of magnitude for both species \citep{Garrod2006, Cazaux2003, Taquet2015}. Our sensitivity is, however, not sufficient to detect dimethyl ether at a similar column density compared to methyl formate.} 

While methyl formate is abundant over the entire extent of the inner envelope, the lowest column density O-bearing COMs, such as ethanol and ethylene glycol are only detected towards the accretion shocks first recognised by the torsionally excited state \methanol\ line in \citetalias{Csengeri2018}. The distribution of these primarily O-bearing COMs is rather similar to the high excitation methanol emission showing two peaks offset from the dust continuum and thus the central protostar (Fig.\,\ref{fig:molecules_all}). Similarly to the high excitation methanol lines, their velocity pattern shows the two velocity components offset by $\sim\pm4.5$\,\kms\ with respect to the \vlsr\ of the source. 
This allows us to conclude that due to the change of physical conditions at the accretion shocks, heavier O-bearing COMs outline well the existence of these shocks.

The N-bearing COMs, especially vinyl and ethyl cyanides, are located towards the immediate vicinity of the protostar {and the accretion disk} and show an extension in the direction of the outflow. This suggests that they are also associated with the innermost region of the outflow cavity, and potentially the accretion disk as also suggested for the vibrationally excited, $\varv_7=1$ state HC$_3$N  transition  at 346.456\, GHz in \citetalias{Csengeri2018}. Our results provide the first direct evidence that, in contrast to O-bearing COMs, molecules with a CN group peak on the innermost few hundred au vicinity of the protostar likely associated with regions of hot gas, which explains why some studies may find them at elevated temperatures relative to the O-bearing COMs \citep{Qin2010, Weaver2017}. 

{In fact models of photodissociation regions (PDRs) show that the CN emission is sensitive to the UV radiation (e.g.\,\citealp{Jansen1995,Sternberg1995,vanZadelhoff2003,Walsh2010}), which could explain the enhancement of molecules with a CN group towards the central position associated with the accretion disk. For example, towards low-mass embedded protostars CN has been proposed to trace the outflow cavity walls \citep{Jorgensen2004}, while CN is also one of the brightest tracers of protoplanetary disks \citep{Guilloteau2014, Cazzoletti2018}, where it has been proposed to trace the UV impacted upper warm molecular layers. } 

{As for hot cores, complex cyanides have been modelled by \citet{Allen2018}, who suggest that a longer warm up time together with a higher cosmic ray ionisation rate is necessary to reproduce the observed high abundance of vinyl and ethyl cyanides.  The chemical differentiation between O- and N-bearing COMs observed toward the Orion Hot Core and Compact Ridge could also be explained with a different thermal evaporation history \citep{Caselli1993}.} 

%A prominent example for chemical differentiation between O- and N-bearing COMs was observed towards the Hot Core and the Compact Ridge in Orion, where models with different thermal evaporation history could reproduce the observed chemical differences \citep{Caselli1993}. 

% CN emission is found to be impacted by UV radiation which could explain the enhancement of these molecules towards the center. Jorgensen et al. 2004

The case of formamide is somewhat in between these two components, as this molecule shows strong emission towards the dust peak while it is also present in the inner envelope, and shows brighter emission towards at least one of the shock positions. The observed difference compared to the other N-bearing COMs may be explained by a different chemical formation pathway due to its amide bond %/peptide link 
($-$N$-$C(=O)$-$), which is also suggested by the fact that it has been detected in various environments, such as low-mass protostars  \citep{Lopez-Sepulcre2015}, also showing a ring around the central object \citep{Coutens2016};  hot cores \citep{Bisschop2007}; %IRAS16293). 
and shocks \citep{Mendoza2014}. 
Although we find a spatial correlation between methanol, methyl formate and formamide, this does not provide enough constraints on the chemical formation routes, that is a grain surface production and sublimation versus gas-phase formation scenario. Therefore, the formation pathway of formamide remains unclear (c.f.\,\citealp{Mendoza2014}), in addition to the gas-phase reactions, its grain surface production could be efficient \citep{Garrod2008}. 

%The distribution of formamide based on our observations suggests that it is present in various components of the envelope, however, despite its strong presence in the inner envelope associated with dust at $T\sim$110\,K \texbf{our observations are not sufficient to conclude on its chemical origin}.

%\subsection{Shock chemistry instead of a radiatively heated envelope}\label{sec:d2}
\subsection{Change of molecular composition at the accretion shocks}\label{sec:d2}

Based on the molecular emission and the kinetic temperatures ($T_{\rm kin}$) derived from our LTE modelling, towards \mysou\ we 
identify three distinct physical components hosting emission of COMs: (1) the inner envelope showing  extended emission of methyl formate, HDO and formamide with $T_{\rm kin}=$110\,K and an extent of $\sim$ 1000-1500\,au FWHM; (2) the shock spots with a higher kinetic temperature of $T_{\rm kin}=$160-190\,K together with an enhanced column density of O-bearing COMs. {Ethanol,  acetone, and ethylene glycol show, for example, the highest column density and abundance} towards the shock positions. {The kinematics of these species further confirms their association with the accretion shocks as they show prominently the two velocity components with $\sim\pm4.5$\,\kms\ offset from the source \vlsr, similarly as the \methanol\ $\varv_{\rm t}=1$ line reported in \citetalias{Csengeri2018}} ; (3) the closest vicinity of the protostar and its accretion disk characterised by N-bearing COMs with a CN group, such as vinyl and ethyl cyanide. %This spatial dichotomy between O- and N-bearing COMs is particularly intriguing, because it is observed using transitions of similar upper energy levels with $E_{\rm up}$=150$-$300\, K (Fig.\,\ref{fig:molecules_all}). 

{Since blending and dust opacity hinders us from a quantitative analysis of the molecular composition towards the central position, here we compare the molecular abundances between the accretion shocks and the inner envelope.} 
Towards the shock positions we estimate a beam averaged H$_2$ column density, $N$(H$_2$), of $1.89-2.04\times10^{24}$\,cm$^{-2}$ assuming a dust temperature, $T_{\rm d}=T_{\rm kin}$ of 180\,K. In contrast, towards the inner envelope we measure a somewhat lower kinetic temperature of 110\,K {and} $N$(H$_2$)=$1.9\times10^{24}$\,cm$^{-2}$. This suggests that considering the temperature variations, the H$_2$ column densities at the three positions are practically identical. Our modelling shows that the abundance of \methanol\ reaches up to {7.8$\times10^{-6}$$-$1.1$\times10^{-5}$ towards the accretion shocks, it is, however, up to a factor of five lower towards the inner envelope with an abundance of 2.2$\times10^{-6}$. The other O-bearing COMs have a molecular abundance relative to H$_2$  between 2.1$\times10^{-9}$ and 5.3$\times10^{-7}$ towards the shock positions, and an order of magnitude lower abundance range, between 3.6$\times10^{-9}$ and 5.4$\times10^{-8}$ towards the position representing the bulk of the inner envelope. Compared to the most abundant O-bearing COMs, such as methyl formate, the N-bearing COMs like vinyl and ethyl cyanide have up to an order of magnitude lower abundance range, between 1.3$\times10^{-8}$ and 5.1$\times10^{-8}$ towards the shock positions. The estimated abundances are also lower towards the inner envelope, and range between 8.6$\times10^{-9}$ and 2.6$\times10^{-8}$ for all the N-bearing COMs. The molecular abundances of formamide are of the order of $10^{-8}$, at least one order of magnitude higher than observed towards hot corino objects \citep{Lopez-Sepulcre2015}, although our estimations are only based on practically one transition and may not be robust for this molecule. The estimated molecular abundances are listed in Table\,\ref{tab:results}. }

% Do the two shock positions show similar molecular composition?
We compare the molecular composition towards the shock positions and the inner envelope in Fig.\,\ref{fig:barplot}. In general, we find that the molecular composition of the two shock spots are broadly consistent with each other, and show very similar molecular abundances for most  species. However, the observed {molecular} abundance of COMs {with respect to H$_2$} towards the inner envelope is {found to be} several factors lower compared to the shock spots. {After methanol,} methyl formate is the second most abundant molecule both in the shock spots and in the envelope. Towards the position of the inner envelope, it shows, however, a much higher relative abundance relative to \methanol. 

{The change of molecular abundances normalised to \methanol\ with respect to the \emph{B} shock position is shown in Fig.\,\ref{fig:barplot2}. This further suggests that the relative change of molecular composition between the \emph{A} and \emph{B} shock spots is small, thus their molecular composition is similar. The molecular composition towards the inner envelope is, however, considerably different, in particular the relative fraction to \methanol\ from methyl formate, ethanol and acetone is smaller in the inner envelope than at the B shock position, suggesting that the gas becomes enriched in these O-bearing COMs in the shocks.} 

{HDO has a similar abundance towards the inner envelope and the accretion shocks, suggesting that not only the accretion shocks but also the inner envelope has a significant contribution to the overall amount of HDO. Its relative fraction compared to methanol changes, however, significantly and becomes considerably higher towards the inner envelope. This is simply because the methanol abundance is lower towards the inner envelope compared to the accretion shocks.}

{In contrast, the cyanides show a relatively moderate change of abundance ratio between the envelope and the accretion shocks.} The C$_2$H$_5$CN/C$_2$H$_3$CN abundance ratio is 2.4$-$3.4 towards all positions, suggesting that saturated nitriles are in general more abundant than unsaturated ones, which is expected for hot cores. % similarly as it is observed for solar system bodies, like Titan \citep{Lai2017}. 
Saturated molecules are chemically more stable which may explain their observed higher abundance. {Chemical models predict that (e.g.\,\citealp{Caselli1993}) C$_2$H$_3$CN forms from C$_2$H$_5$CN through ion-molecule reactions in the gas phase, therefore their ratio has been suggested as a tracer for the chemical age of hot cores. For example \citet{Fontani2007} find C$_2$H$_5$CN/C$_2$H$_3$CN abundance ratios around 2.0$-$3.3 towards six classical hot cores and conclude that their age is less than 10$^5$ years. Such a ratio is also consistent with that found towards Orion-KL and SgrB2(N) \citep{Zeng2018}, however towards SgrB2(N) high angular resolution measurements resolving the individual hot cores suggest very different ratios and find C$_2$H$_5$CN/C$_2$H$_3$CN=5$-$17. The evolutionary sequence towards SgrB2(N) based on the  C$_2$H$_5$CN/C$_2$H$_3$CN ratio is, however, contradicting the results of dedicated physico-chemical modelling \citep{Bonfand2019} suggesting that either the chemical network lacks some reactions or that other physical parameters such as the stellar mass and ionisation rate may have a stronger impact on the chemical evolution than the age. }

   \begin{figure}[!h]
%   \resizebox{\hsize}{!}
            {\includegraphics[width=0.9\linewidth]{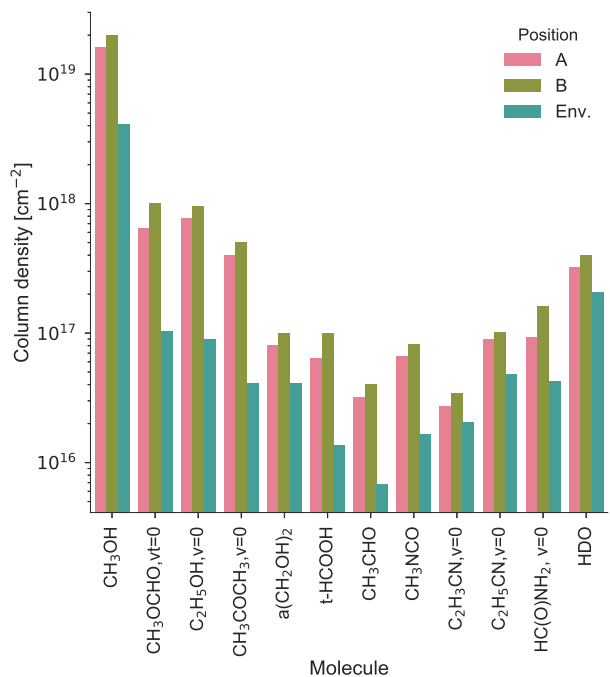}}
      \caption{Molecular column densities obtained from the Weeds LTE modelling of COMs and HDO. The three positions (shock-\textsl{A}, -\textsl{B} and the position toward the inner envelope) are labelled in different colours.}
         \label{fig:barplot}
   \end{figure}
%
%-------------------------------------------------------------
   \begin{figure}[!h]
%   \resizebox{\hsize}{!}
            {\includegraphics[width=0.9\linewidth]{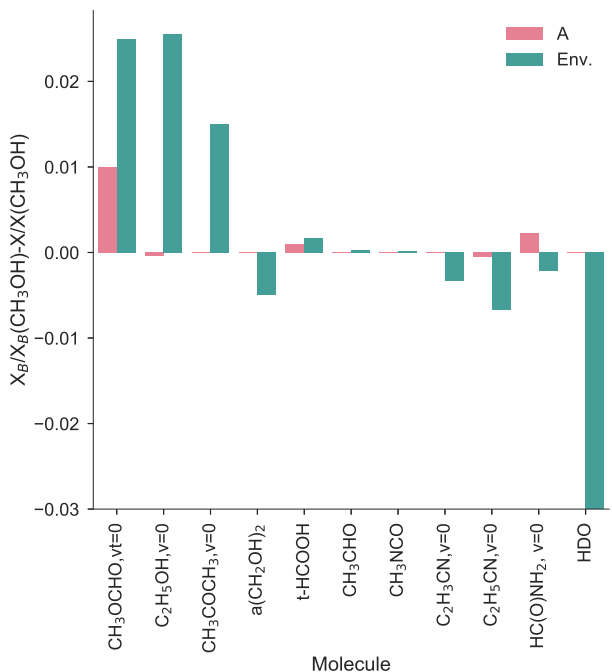}}
      \caption{Relative molecular abundances with respect to that of the \textsl{B}-shock position. Positive means a relative decrease  of certain species compared to the same molecular abundance derived at the \textsl{B}-shock position. }
         \label{fig:barplot2}
   \end{figure}
%
%-------------------------------------------------------------

%\subsection{Origin of the different chemistry?}

{Altogether this quantitatively demonstrates a change in the molecular composition of COMs between the envelope and the accretion shocks, while Fig.\,\ref{fig:molecules_all} shows that also the central position is likely to have an even more drastic change of molecular composition. The origin of this different chemistry observed towards the central position, the accretion shocks and the inner envelope needs to be understood. 
As discussed earlier, the chemistry of N-bearing COMs with a CN group may be influenced by a strong change of physical conditions in the immediate vicinity of the protostar, such as UV radiation as well as the gas being exposed to a heating source for a longer time-scale compared to the gas within the cold envelope. 
A different thermal history of the gas, together with different initial conditions for the chemistry in the envelope due to thermal or non-thermal desorption processes could be responsible for the observed spatial segregation of molecules.}

\subsection{Distribution of deuterated water}\label{sec:d4}

HDO from cold and hot gas has been detected towards classical hot cores both at low and high angular resolution (e.g.\,\citealp{Jacq1990, vanderTak2006, Liu2013}).
Here we resolve for the first time HDO emission towards the inner regions of a high-mass protostellar envelope %that we propose to correspond to a hot core precursor. 
{and find that it is associated with the inner envelope.
Its velocity field  is axisymmetric roughly perpendicular to the outflow direction consistent with a rotational pattern also observed in several \methanol\ transitions. A similar rotational pattern has also been observed towards AFGL 2591, where, based on its correlation with the continuum structure, it has been proposed to trace a circumstellar disk \citep{vanderTak2006}.
In addition, HDO shows an enhanced molecular column density towards at least one of the shock positions, and similarly to the O-bearing COMs, its emission does not show a prominent peak on the position of the protostar. }
 
Since we only have one observed transition, the estimated parameters of HDO rely on the assumption that the emitting gas has the same rotational temperature as  \methanol, which is 180\,K for the shock position and 110\,K on the position of the inner envelope. 
This gives an estimate of the HDO column density of $2.1$ and $4.0$$\times$10$^{17}$\,cm$^{-2}$ towards the inner envelope and the shock position, respectively. The HDO column density is relatively similar towards the shock spots and the inner envelope, however its relative fraction compared to \methanol\ is higher in the envelope than in the accretion shocks by a factor of 2.  As a comparison, towards G34.26+0.15, one of the classical hot cores, \citet{Coutens2014} estimate an HDO column density of $\sim$1.6$\times10^{16}$\,cm$^{-2}$ with an excitation temperature of $\sim$79\,K towards the inner regions of the hot core. %, while low angular resolution observations towards other high-mass star forming regions by Liu et al. (in prep) find up to an order of two lower column densities for HDO.
Our estimates give an HDO abundance of 
$X$(HDO)=$1.1-2.1\times10^{-7}$ relative to H$_2$, which is of the same order of magnitude as \citet{Liu2011} find for the T$>$100K regime for G34.26+0.15, and what \citet{Kulczak2017} estimates for the inner region of other classical hot cores. 

{The kinetic temperature  of the inner envelope is mainly estimated based on methanol and methyl formate, and because HDO shows a similar spatial distribution to these species, it is reasonable to assume that it is also associated with gas close to, or just above the ice sublimation temperature, $T=90-190$\,K.} {This would suggest that} the observed HDO emission likely originates from the evaporation of the grain surfaces, expected at $\gtrsim$100\,K, \citep{Fraser2001}. This is also consistent with the abundance jump assumed by modelling of single dish (low resolution) observations for classical hot cores, as well as low- to intermediate mass Class 0 protostars \citep{Comito2003, Parise2005a, Liu2013}.  %The proposed high deuteration fractionation supports this picture, together with the moderately warm gas temperatures of $T=90-190$\,K observed at the shock spots and the inner envelope. %Our observations suggest an efficient mechanism for water deuteration on the grain surfaces.

{Our detection of HDO from the inner envelope is intriguing, because} 
chemical models predict a decrease in the abundance of water, and therefore HDO towards the inner 1000\,au of hot cores \citep{Coutens2014}, with the destruction of HDO being even more efficient than the destruction of H$_2$O. In comparison to our results, both the location and the relatively high abundance of HDO suggest that H$_2$O, and HDO destruction has not been efficient in this source yet, {providing further evidence that the source is chemically young.}

\section{Conclusions}\label{sec:conclusions}

Using high angular-resolution observations with ALMA, we investigate the physical and chemical structure of the massive envelope of a high-mass protostar, \mysou. We identify emission from 10 COMs using LTE modelling of a 7.5\,GHz non continuous spectral coverage around 345\,GHz, and find that its overall molecular composition shows a similar richness compared to other high-mass star forming regions. 

Comparing to recent observations of hot cores, the radiatively heated inner region is the most compact observed to date with a radius ($R_{\rm 90\%}$) of $<900$\,au estimated from vinyl and ethyl cyanides. Inferred from \methanol\ transitions, the bulk of the gas in the inner envelope is at a temperature of $T_{\rm kin}=110$\,K, allowing us to witness the emergence of accretion shocks suggesting a different heating mechanism compared to the classical, radiatively heated hot cores. However, given the $\sim$50\,M$_{\odot}$ expected final stellar mass, \mysou\ is likely to represent an earlier evolutionary stage compared to classical hot cores.

For the first time, we spatially resolve on $<$1000\,au scales within a single collapsing envelope the chemical differentiation of O-bearing COMs versus N-bearing COMs, in particular those with a CN group. The O-bearing COMs, such as ethanol, acetone, ethylene glycol and acetaldehyde have an increased abundance towards the two localised positions of accretion shocks with $T_{\rm kin}$=180\,K. These transitions show the same velocity pattern as the \methanol\ $\varv_t=1$ emission at 334.436\,GHz reported in \citetalias{Csengeri2018}, representing the accretion shocks. 
Emission from other COMs, such as methanol, methyl formate, and formamide is found towards an extended region with a radius ($R_{\rm 90\%}$) of 1175$-$1450\,au and $T_{\rm kin}$=110\,K. %The overall molecular composition of this source is dominated by molecules originating from the accretion shocks instead of the radiatively heated inner region corresponding to the emerging hot core. 

We image and spatially resolve emission from the HDO ($J=3_{3,1}-4_{2,2}$) line. It is extended over the inner envelope and its velocity pattern is consistent with a velocity gradient roughly perpendicular to the outflow, likely corresponding to the rotational pattern of the inner envelope. Its extent is comparable to that of methyl formate with $T_{\rm kin}$=110\,K. The high HDO column density {suggests that the destruction of water has not been efficient towards this source since HDO is expected to be destroyed in the inner regions of classical hot cores.} %a high deuterium fractionation, which, if a result of grain surface processes, is expected to lead to a higher abundance of deuterated COMs.

We identify three physical components within the envelope, a compact inner region representing the immediate vicinity of the protostar and its accretion disk, the inner envelope, and the accretion shocks within the envelope. We quantitatively show that the molecular composition of COMs towards the two accretion shocks is similar, while there is a change of molecular composition among COMs compared to the inner envelope. In addition, we qualitatively show that the central regions are particularly apparent in N-bearing COMs with a CN group, such as vinyl and ethyl cyanide that could explain the typically observed higher temperatures found with these molecules towards classical hot cores.

%The physical conditions of the emitting gas and the estimated molecular abundances of various COMs suggest that accretion shocks in the infalling envelope may have a change in their molecular composition, leading to higher abundances of various, mostly, O-bearing COMs. The N-bearing COMs with a CN group show a different behaviour compared to other COMs, and pinpoint either different physical conditions of the emitting gas, or a change in molecular composition towards the innermost regions of the envelope at the immediate vicinity of the protostellar embryo.

\begin{acknowledgements}
This paper makes use of the ALMA data: ADS/JAO.ALMA 2013.1.00960.S. ALMA is a partnership of ESO (representing its member states), NSF (USA), and NINS (Japan), together with NRC (Canada), NSC and ASIAA (Taiwan), and KASI (Republic of Korea), in cooperation with the Republic of Chile. The Joint ALMA Observatory is operated by ESO, AUI/NRAO, and NAOJ. T.Cs. acknowledges support from the \emph{Deut\-sche For\-schungs\-ge\-mein\-schaft, DFG\/}  via the SPP (priority programme) 1573 'Physics of the ISM'. 
\end{acknowledgements}

   \bibliographystyle{aa} % style aa.bst
   \bibliography{bib} % your references Yourfile.bib

\begin{appendix} %First appendix
\section{Spectra obtained towards the \emph{A}-shock position, and the inner envelope.}\label{app:other_positions}

%-------------------------------------------------------------
%                                             Two column Figure 
%-------------------------------------------------------------
   \begin{sidewaysfigure*}
%   \begin{figure*}
   \centering
            {\includegraphics[width=0.48\linewidth]{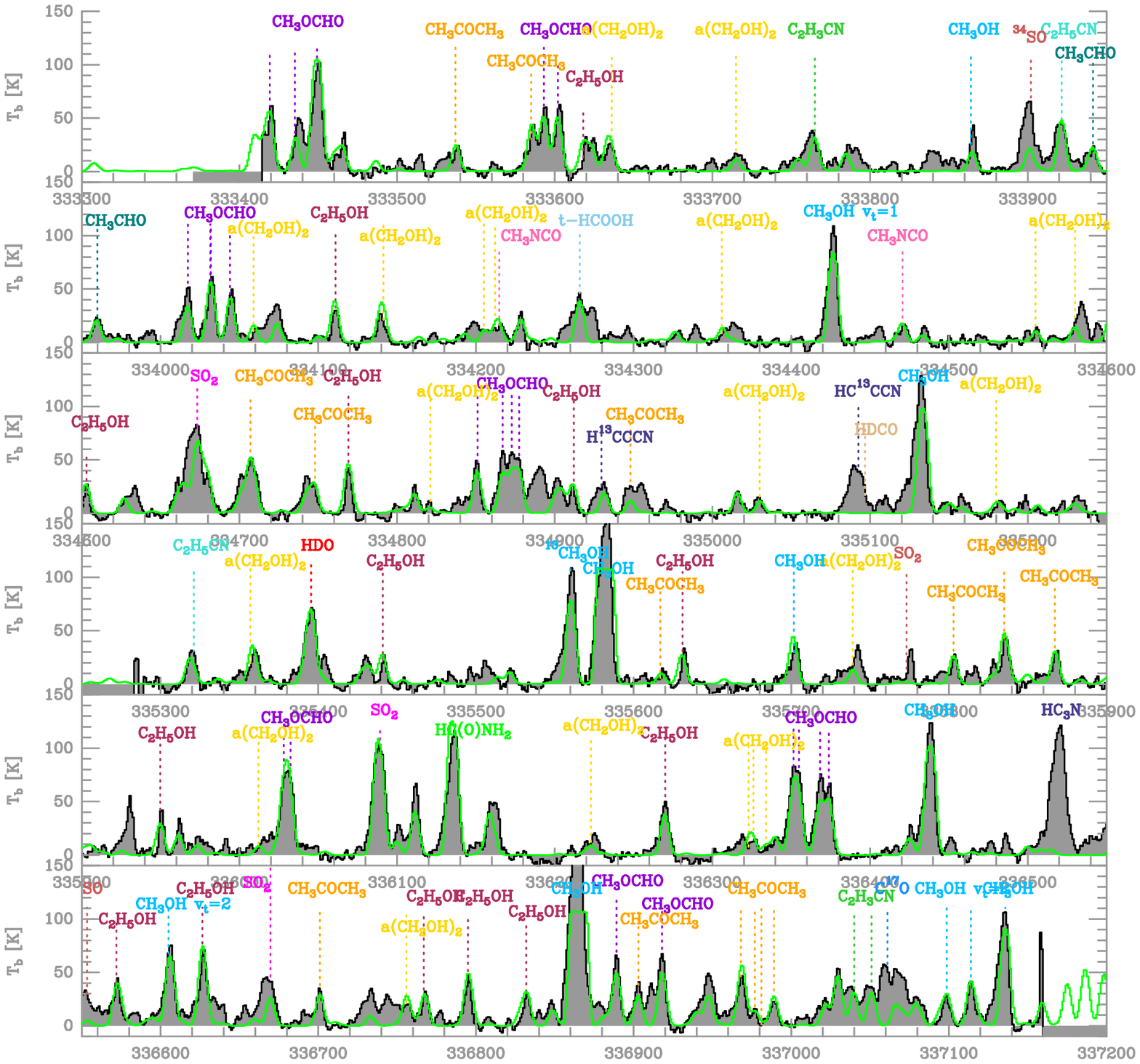}}
         {\includegraphics[width=0.48\linewidth]{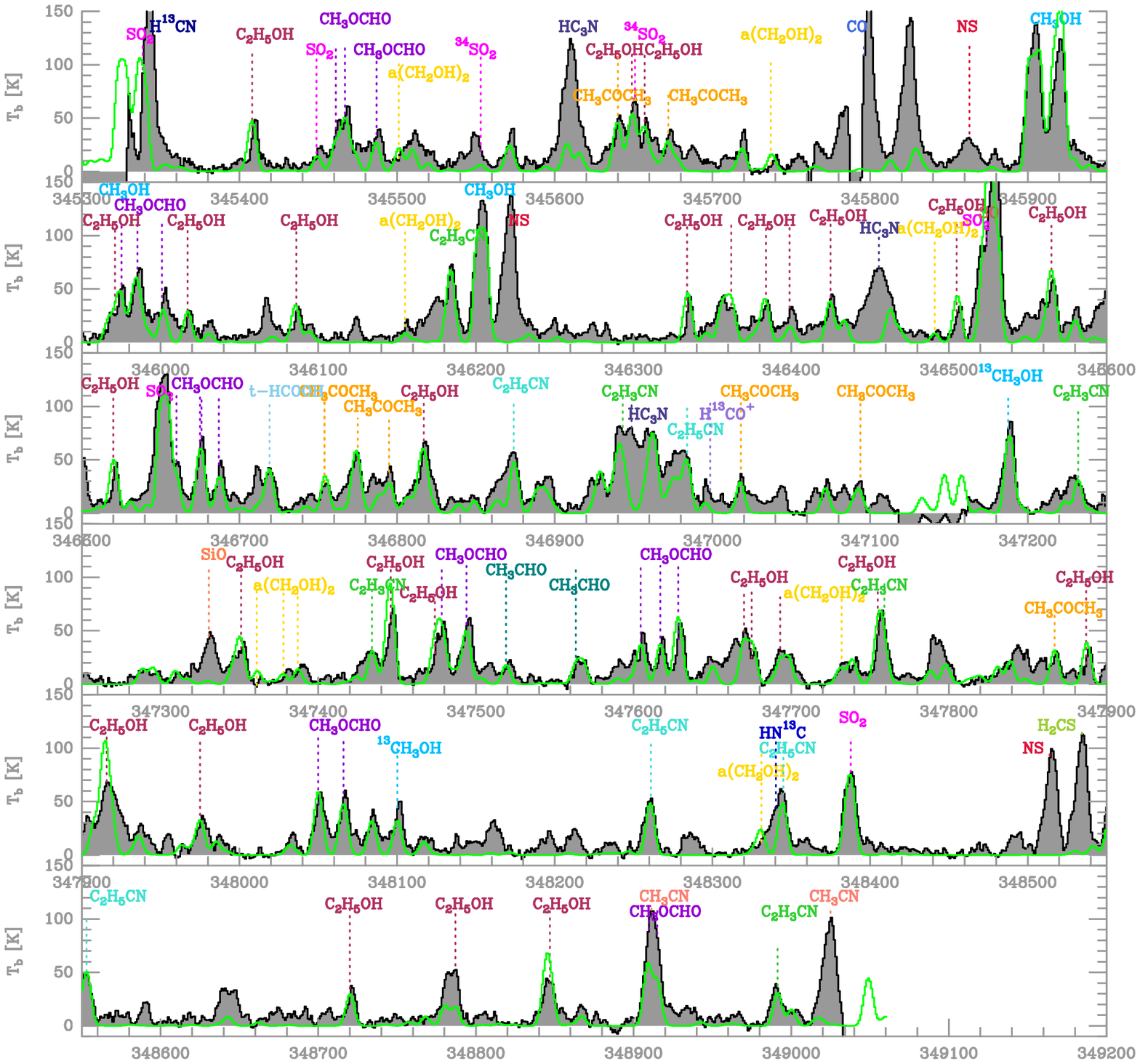}}
      \caption{Same as Fig.\,\ref{fig:spectra} extracted towards the \textsl{A}-shock position.   }
         \label{fig:spectra_v3} % envelope p1
%   \end{figure*}
   \end{sidewaysfigure*}
 \begin{sidewaysfigure*}
%   \begin{figure*}
   \centering
            {\includegraphics[width=0.48\linewidth]{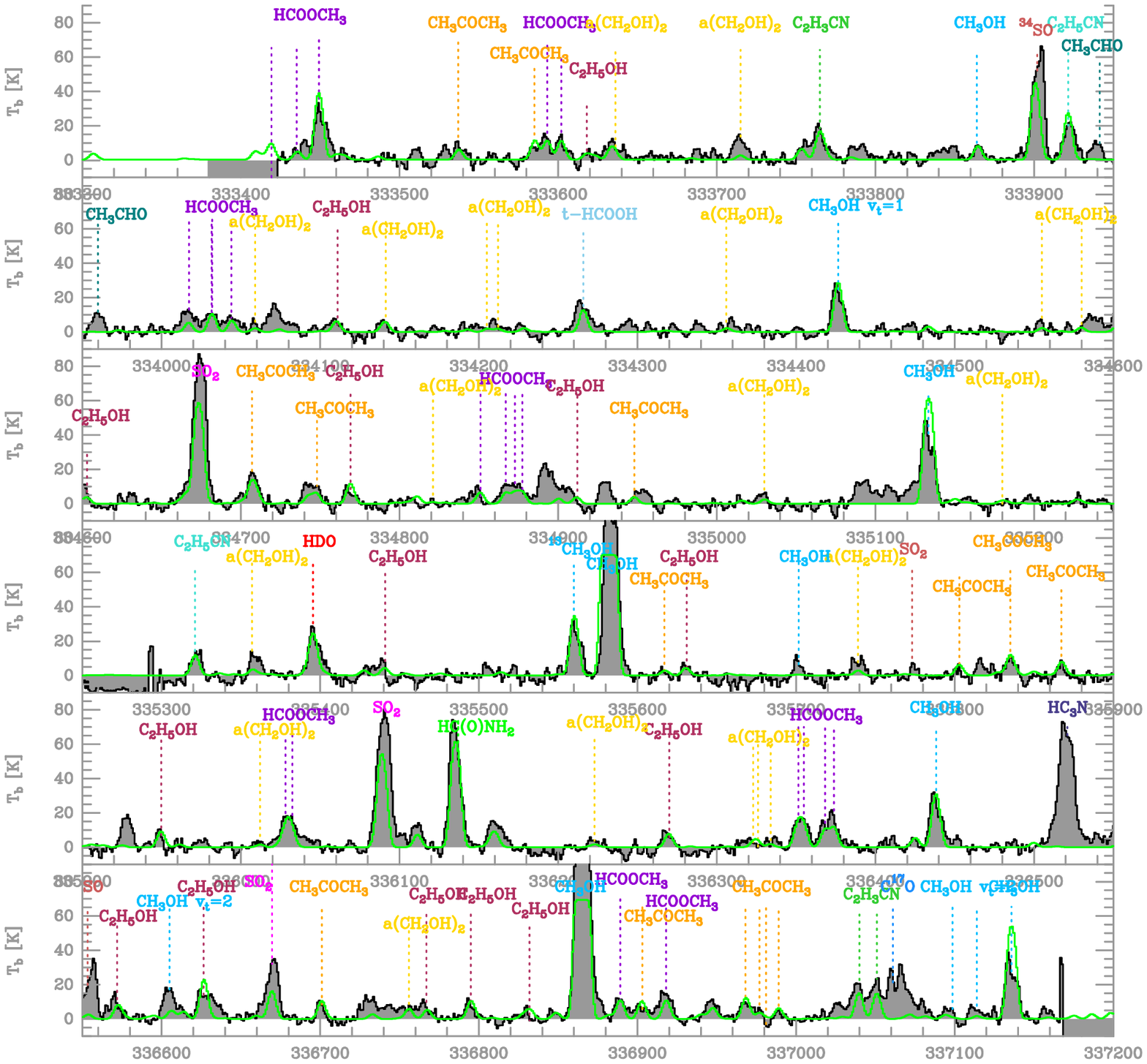}}
         {\includegraphics[width=0.48\linewidth]{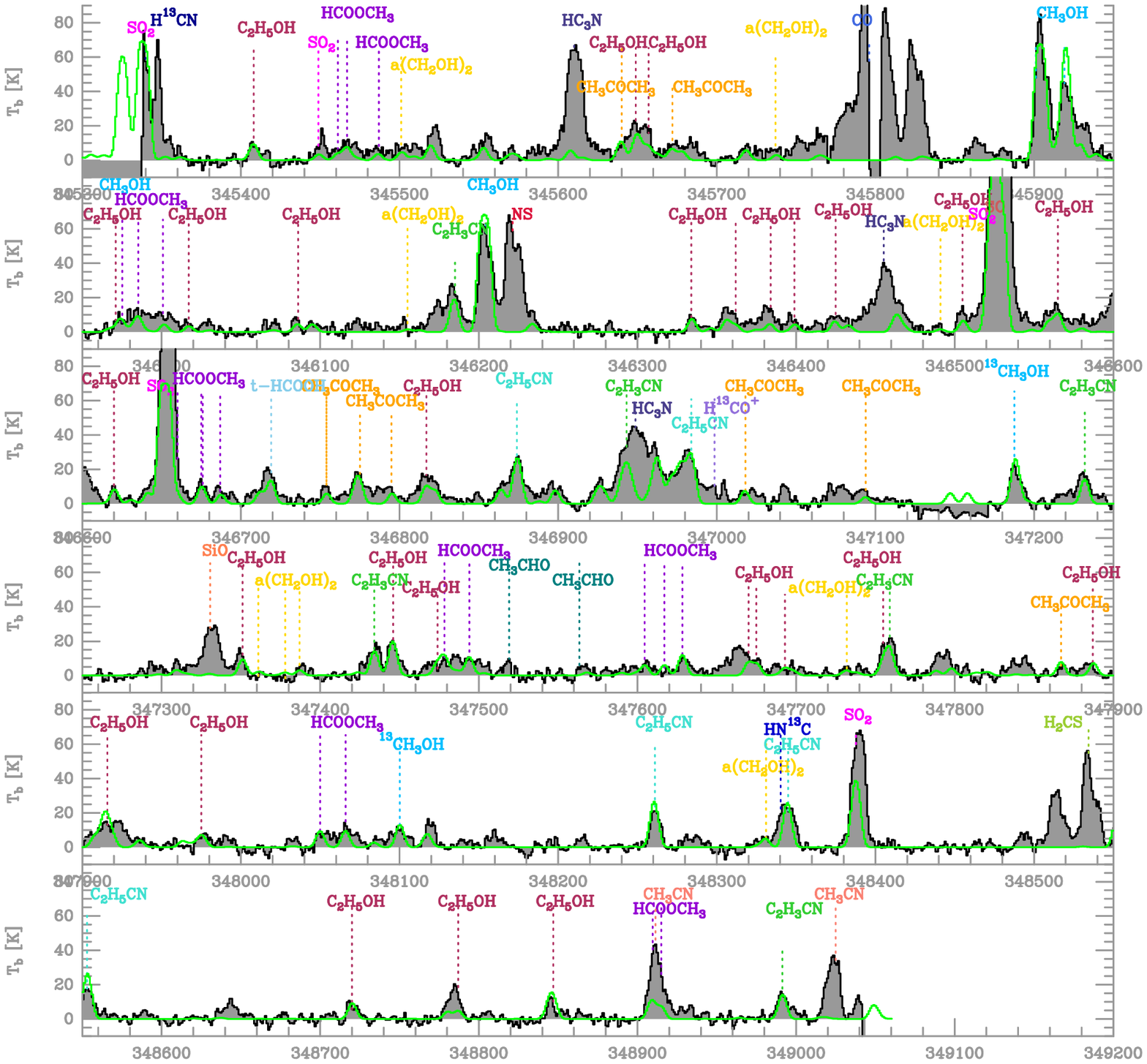}}
      \caption{Same as Fig.\,\ref{fig:spectra} extracted towards a position of the inner envelope.   }
         \label{fig:spectra_v3} % envelope p1
%   \end{figure*}
   \end{sidewaysfigure*}
%-------------------------------------------------------------

%\section{Summary of fits to the individual transitions}\label{app:lists}
%\subsection{Methyl formate}
%\input{linelist_methyl_formate}
%\input{figures_lines}

%\subsection{Ethanol}
%\input{linelist_ethanol}
%\input{figures_lines_ethanol}
 
%\subsection{Acetone}
%\input{linelist_acetone}
%\input{figures_lines_acetone}

%\subsection{Ethylene glycol}
%\input{linelist_ethanol}
%\input{figures_lines_eglycol}
%
%\subsection{Formic acid}
%\input{linelist_ethanol}
%\input{figures_lines_formic_acid}

%\subsection{Vinyl cyanide}
%\input{linelist_ethanol}
%\input{figures_lines_vinyl_cyanide}

%\section{Modeling}
%-------------------------------------- Two column figure (place early!)

\end{appendix}

\end{document}